\documentclass[12pt]{article}
\usepackage[english]{babel}
\usepackage[latin1]{inputenc}
\usepackage{amsfonts,amssymb,amsmath, epsfig}
\usepackage{color,graphicx,graphics,psfrag}
\usepackage{amsmath,amstext,amssymb,amsfonts, amscd}
\usepackage{amsthm}
\usepackage{vmargin} 
\usepackage[only,llbracket,rrbracket]{stmaryrd}

\usepackage{hyperref}

\def\Box{\leavevmode\vbox{\hrule
     \hbox{\vrule\kern4pt\vbox{\kern4pt}%
           \vrule}\hrule}}
\def\blackbox{\leavevmode\vrule height 5pt width 4pt depth 0pt\relax}
\def\endproof{\null\hfill {$\blackbox$}\bigskip}

\newcounter{appendix}
\def\appendix{\advance\c@appendix by 1
   \def\thesection{\Alph{section}}
   \ifnum\c@appendix=1 \setcounter{section}{-1} \fi
   \@startsection {section}{1}{\z@}{-3.5ex plus -1ex minus 
   -.2ex}{2.3ex plus .2ex}{\Large\bf}}

\def\paragraph#1{{\bf #1\ }}

\newtheorem{lemma}{Lemma}[section]  

\newtheorem{theorem}[lemma]{Theorem}

\newtheorem{definition}[lemma]{Definition}

\newtheorem{proposition}[lemma]{Proposition}

\newtheorem{remark}{Remark}[section]

\newtheorem{conjecture}{Conjecture}[section]

\title{Macroscopic limits and phase transition in a system of self-propelled particles}

\author{Pierre Degond$^{(1,2)}$, Amic Frouvelle$^{(1,2)}$,  Jian-Guo Liu$^{(3)}$}
\date{}
\begin{document}
\maketitle

\vspace{0.5 cm}

\begin{center}
1-Université de Toulouse; UPS, INSA, UT1, UTM ;\\ 
Institut de Mathématiques de Toulouse ; \\
F-31062 Toulouse, France. \\
2-CNRS; Institut de Mathématiques de Toulouse UMR 5219 ;\\ 
F-31062 Toulouse, France.\\
email: pierre.degond@math.univ-toulouse.fr, amic.frouvelle@math.univ-toulouse.fr
\end{center}

\begin{center}
3- Department of Physics and Department of Mathematics\\
Duke University\\
Durham, NC 27708, USA\\
email: jliu@phy.duke.edu
\end{center}

\begin{abstract}
We investigate systems of self-propelled particles with alignment interaction. Compared to previous work \cite{degond2008continuum, frouvelle2011continuum}, the force acting on the particles is not normalized and this modification gives rise to phase transitions from disordered states at low density to aligned states at high densities. This model is the space inhomogeneous extension of \cite{frouvelle2011dynamics} in which the existence and stability of the equilibrium states were investigated. When the density is lower than a threshold value, the dynamics is described by a non-linear diffusion equation. By contrast, when the density is larger than this threshold value, the dynamics is described by a similar hydrodynamic model for self-alignment interactions as derived in \cite{degond2008continuum, frouvelle2011continuum}. However, the modified normalization of the force gives rise to different convection speeds and the resulting model may lose its hyperbolicity in some regions of the state space.
\end{abstract}

\medskip
\noindent
{\bf Acknowledgements:} The authors wish to acknowledge the hospitality of Mathematical Sciences Center and Mathematics Department of Tsinghua University where this research was completed. The research of J.-G. L. was partially supported by NSF grant DMS 10-11738.

\medskip
\textbf{Key words:} self-propelled particles, alignment interaction, Vicsek model, phase transition, hydrodynamic limit, diffusion limit, Chapman-Enskog expansion, non-hyperbolicity.

\medskip
\textbf{AMS subject classification:} 35L60, 35K55, 35Q80, 82C05, 82C22, 82C70, 92D50.

\setcounter{equation}{0}
\section{Introduction}
\label{sec:into}

The context of this paper is that of \cite{degond2008continuum} and is concerned with a kinetic model for self-propelled particles and its hydrodynamic or diffusion limits. The particles move with the same constant speed and their velocity directions  (which belong to the sphere~${\mathbb S}$) align to the local average orientation, up to the addition of some noise. This model has been proposed as a variant of the Vicsek particle model \cite{Vicsek}. In this paper, we remove the normalization of the force intensity which was done in \cite{degond2008continuum}. This apparently minor modification leads to the appearance of phase transitions, which have been studied in the space-homogeneous setting in \cite{frouvelle2011dynamics}. In \cite{frouvelle2011dynamics}, it is proved that the equilibrium distribution function changes type according to whether the density is below or above a certain threshold value. Below this value, the only equilibrium distribution is isotropic in velocity direction and is stable. Any initial distribution relaxes exponentially fast to this isotropic equilibrium state. By contrast, when the density is above the threshold, a second class of anisotropic equilibria formed by Von-Mises-Fischer distributions of arbitrary orientation appears. The isotropic equilibria become unstable and any initial distribution relaxes towards one of these anisotropic states with exponential speed of convergence. 
We would like to emphasize the connection of the presented alignment models to the the Doi-Onsager \cite{DE, Onsager} and Maier-Saupe \cite{MS} models for phase transition in polymers. The occurrence of phase transitions makes a strong difference in the resulting macroscopic models as compared with the ones found in \cite{degond2008continuum, frouvelle2011dynamics}, where no such phase transitions were present. 

In the present paper, we rely on this previous analysis to study the large-scale limit of the space-inhomogeneous system. In the regions where the density is below the threshold, the convection speed becomes zero and the large-scale dynamics becomes a nonlinear diffusion. On the other hand, in the region where the density is above the threshold, the large-scale dynamics is described by a similar hydrodynamic model for self-alignment interactions as derived in \cite{degond2008continuum, frouvelle2011continuum}. However, the modified normalization of the force gives rise to different convection speeds and the resulting model may lose its hyperbolicity in some regions of the state space.

The Vicsek model \cite{Vicsek}, among other phenomena, models the behaviour of individuals in animal groups such as fish schools, bird flocks, herds of mammalians, etc (see also \cite{Aldana_Huepe, Aoki, Couzin, Gregoire_Chate}). This particle model (also called 'Individual-Based Model' or 'Agent-Based model') consists of a discrete stochastic system for the particle positions and velocities. A time-continuous version of the Vicsek model and its kinetic formulation have been proposed in \cite{degond2008continuum}.  The rigorous derivation of this kinetic model has been performed in \cite{bolley2011meanfield}. 

Hydrodynamic models are more efficient than particle models for large numbers of particles, because they simply encode the different particles quantities into simple averages, such as the density or mean-velocity. 
We refer to \cite{Chuang, Dorsogna, Mogilner1, Mogilner2, toner1998flocks, TB1, TB2} for other models of self-propelled particle interactions. Rigorous derivations of hydrodynamic models from kinetic ones for self-propelled particles are scarce and \cite{degond2008continuum, frouvelle2011continuum} are among the first ones (see also some phenomenological derivations in \cite{KRZB,RBKZ,RKZB}). Similar models have also been found in relation to the so-called Persistent Turning Walker model of fish behavior \cite{DM0, degond2010macroscopic}. Diffusive corrections have also been computed  in \cite{degond2010diffusion}. We refer to \cite{CDP, CKMT} for other macroscopic models of swarming particle systems derived from kinetic theory. In particular, we mention \cite{frouvelle2011continuum} where a vision angle and the dependence of alignment frequency upon local density have been investigated.

The outline of this paper is as follows. In Section~\ref{particular-mean-field}, we describe the Individual-Based  Model (IBM), and its mean-field limit. In Section~\ref{macroscopic-limit}, we investigate the properties of the rescaled mean-field model. We prove that there are two possibilities for a local equilibrium, depending on the value of its density~$\rho$.
Section~\ref{section-disorder} is devoted to the derivation of the diffusion model when the density~$\rho$ is below the threshold. Finally, in Section~\ref{section-order}, we derive the hydrodynamic model for self-alignment interactions in the region where~$\rho$ is above the threshold and study its hyperbolicity. A conclusion is drawn in section \ref{sec:conclu}. Two appendices are added. In appendix 1, we calculate a Poincar\'e constant which provides us with a fine estimate of the rate of convergence to the equilibrium states. In appendix 2. some numerical computations of the coefficients of the model are given.

\setcounter{equation}{0}
\section{Particle system and mean-field limit}
\label{particular-mean-field}

We consider~$N$ oriented particles in~$\mathbb{R}^n$, described by their positions~$X_1,\dots X_N$ and their orientation vectors~$\omega_1,\dots,\omega_N$ belonging to~$\mathbb{S}$, the unit sphere of~$\mathbb{R}^n$. We define the mean momentum~$J_k$ of the neighbors of the particle~$k$ by
\begin{equation*}
J_k = \frac1N\sum_{j=1}^N K(X_j-X_k) \omega_j. 
\end{equation*}
In this paper, the observation kernel~$K$ will be supposed isotropic (depending only on the distance~$|X_j-X_k|$ between the particle and its neighbors), smooth and with compact support. 
Introducing a non-isotropic observation kernel, as in~\cite{frouvelle2011continuum} would lead to the same conclusion, with a slightly different convection speed for the orientation in the macroscopic model, but the computations are more complicated. Therefore we focus on an isotropic observation kernel for the sake of simplicity.

The particles satisfy the following system of coupled stochastic differential equations (which must be understood in the Stratonovich sense), for~$k\in\llbracket1,N\rrbracket$:
\begin{align}
\mathrm d X_k&= \omega_k\,\mathrm d t\label{d_X_k}\\
\mathrm d \omega_k &=  (\mbox{Id} - \omega_k \otimes \omega_k) J_k \, \mathrm d t + \sqrt{2d} (\mbox{Id} - \omega_k \otimes \omega_k)\,\circ \mathrm d B^k_t, 
\label{d_omega_k} 
\end{align}
The first equation expresses the fact that particles move at constant speed equal to unity, following their orientation~$\omega_k$.
The terms~$B^k_t$ stand for~$N$ independent standard Brownian motions on~$\mathbb{R}^n$, and the projection term~$(\mbox{Id} - \omega_k \otimes \omega_k)$ (projection orthogonally to~$\omega_k$) constrains the norm of~$\omega_k$ to be~$1$. 
We have that~$(\mbox{Id} - \omega_k \otimes \omega_k) J_k=\nabla_\omega(\omega\cdot J_k)|_{\omega=\omega_k}$, where~$\nabla_\omega$ is the tangential gradient on the sphere.
So the second equation can be understood as a relaxation (with a rate proportional to the norm of~$J_k$) towards a unit vector in the direction of~$J_k$, subjected to a Brownian motion on the sphere with intensity~$\sqrt{2d}$. 
We refer to \cite{hsu2002stochastic} for more details on Brownian motions on Riemannian manifolds. 

The interaction term (first term of (\ref{d_omega_k})) is the sum of smooth binary interactions. This model is an intermediate between the Cucker-Smale model~\cite{cucker2007emergent}, where there is no constraint on the velocity and no noise, and the time-continuous version of the Vicsek model proposed in \cite{degond2008continuum}, where the velocity is constant and noise is added. Indeed, in \cite{degond2008continuum},~$J_k$ is replaced by~$\nu \Omega_k$, where~$\Omega_k=\frac{J_k}{|J_k|}$ is the unit vector in the direction of~$J_k$ and the relaxation frequency~$\nu$ is a constant. Therefore, in \cite{degond2008continuum}, the interaction term cannot be recast as a sum of binary interactions and has a singularity when~$J_k$ is close to~$0$. The model presented here brings a modification consisting in letting~$\nu$ depend (linearly) on the norm of the velocity~$J_k$. A related modification has previously been introduced in~\cite{frouvelle2011continuum}, consisting in letting the relaxation parameter~$\nu$ depend on a local density~$\bar{\rho}_k$, but the modification considered here brings newer phase transition phenomena. 

From the Individual-Based Model (\ref{d_X_k}), (\ref{d_omega_k}), we derive a mean-field limit as the number of particles~$N$ tends to infinity.
We define the empirical distribution~$f^N$ by
\begin{equation*}
f^N(x,\omega,t)=\frac1N\sum_{i=1}^N \delta_{(X_i(t),\omega_i(t))}(x,\omega),
\end{equation*}
where the Dirac distribution is defined by duality by~$\langle\delta_{(X,\Omega)},\varphi\rangle_{\mathbb{R}^n\times\mathbb{S}}=\varphi(X,\Omega)$ for any smooth function~$\varphi\in C(\mathbb{R}^n\times\mathbb{S})$, the duality product~$\langle\cdot,\cdot\rangle_{\mathbb{R}^n\times\mathbb{S}}$ extending the usual inner product of~$L^2(\mathbb{R}^n\times\mathbb{S})$. 
For convenience, the integration measure is supposed of total mass equal to~$1$ on the sphere~$\mathbb{S}$, and we have~$\langle f^N,1\rangle_{\mathbb{R}^n\times\mathbb{S}}=1$. 
Denoting the convolution with respect to the space variable by~$*$, and the duality product on the sphere by~$\langle\cdot,\cdot\rangle_{\mathbb{S}}$, we get~$J_k=\langle K*f^N(X_k),\omega\rangle_\mathbb{S}$. If there is no noise ($d=0$), it is easy to see that~$f^N$ satisfies the following partial differential equation (in the sense of distributions): 
\begin{equation*}
\partial_t f^N + \omega\cdot\nabla_xf^N + \nabla_\omega \cdot ((\mbox{Id} - \omega \otimes \omega) \bar{J}_{f^N} f^N)=0,
\end{equation*}
where~$\nabla_\omega\cdot$ denotes the divergence operator on the unit sphere, and
\begin{equation*}
\bar{J}_{f^N}(x,t) = \langle(K*f^N)(x),\omega\rangle_\mathbb{S}. 
\end{equation*}
When noise is present ($d \not = 0$), the empirical distribution~$f^N$ tends to a probability density function~$f$ satisfying the following partial differential equation:
\begin{equation}
\partial_t f + \omega\cdot\nabla_xf + \nabla_\omega \cdot ((\mbox{Id} - \omega \otimes \omega) \bar{J}_f f)=d \Delta_\omega f,
\label{KFP_mf}
\end{equation}
with
\begin{equation}
\bar{J}_f(x,t) = \int_\mathbb{S} (K*f)(x,\omega,t) \,\omega \, \mathrm d\omega.
\label{def_J_bar}
\end{equation}
This result has been shown in \cite{bolley2011meanfield}, under the assumption that the kernel~$K$ is Lipschitz and bounded.

Eqs. (\ref{KFP_mf}), (\ref{def_J_bar}) are the starting point of our study. 
We notice that there is a competition between the alignment and diffusion terms. The alignment term is quadratic while the diffusion term is linear.
So we can expect that alignment wins over diffusion for high densities while at low densities, diffusion dominates. This is the source of the phase transition rigorously studied in the space-homogeneous setting in \cite{frouvelle2011dynamics}. In this reference, it is proven that there is a unique isotropic equilibrium at low density but beyond a certain density threshold, another family of non-isotropic equilibria in the form of Von-Mises-Fischer distributions arises. Above this threshold, the isotropic equilibria become unstable and the anisotropic ones become the stable ones. Therefore, we expect different large-scale limits according to whether the density is lower or larger than this threshold. 

We now make some preliminary remarks and assumptions. We suppose that the kernel~$K$ is integrable, and that its total weight~$K_0=\int_{\mathbb{R}^n}K(x)\mathrm d x$ is positive. Writing
\begin{equation*}
\widetilde{f}(x,\omega,t)= f(\tfrac1dx,\omega,\tfrac1dt)\quad \text{ and }\quad  \widetilde{K}(x)=\frac1{K_0 d^n}K(\tfrac1dx),
\end{equation*}
we get that~$\widetilde{f}$ satisfies \eqref{KFP_mf} with~$d=1$ and~$K$ replaced by~$\widetilde{K}$ in~\eqref{def_J_bar}, and we have~
$$ \int_{\mathbb{R}^n}\widetilde{K}(x)\mathrm d x=1.$$ 
So without loss of generality, we can suppose that~$d=1$ and that~$K_0=1$.

We are now ready to investigate the large-scale behavior of (\ref{KFP_mf}), (\ref{def_J_bar}) in space and time. The derivation of the macroscopic limit proceeds as in~\cite{degond2008continuum}, and follows closely the presentation of~\cite{frouvelle2011continuum}, so we only give a summary, focusing on the points which are specific to the present model, in particular the distinction between the ordered and disordered phases.

\setcounter{equation}{0}
\section{The macroscopic limit}
\label{macroscopic-limit}

\subsection{Hydrodynamic scaling}
\label{subsec:hydro}

In order to observe the system at large scales, we perform a hydrodynamic scaling. 
We introduce a small parameter~$\varepsilon$, and the change of variables~$x'=\varepsilon x$,~$t'=\varepsilon t$.
We write~$f^\varepsilon(x',\omega,t')=f(x,\omega,t)$, and ~$K^\varepsilon(x')=\frac1{\varepsilon^n}K(x)$.
Then~$f^\varepsilon$ satisfies
\begin{equation}
\label{KFP_eps} 
\varepsilon(\partial_t f^\varepsilon + \omega\cdot\nabla_xf^\varepsilon) = - \nabla_\omega \cdot ((\mbox{Id} - \omega \otimes \omega) \bar{J}^\varepsilon_{f^\varepsilon} f^\varepsilon) +  \Delta_\omega f^\varepsilon ,   
\end{equation}
with
\begin{equation}
\bar{J}^\varepsilon_{f^\varepsilon}(x,t) = \int_{\mathbb{S}} (K^\varepsilon*f^\varepsilon)(x,\omega,t) \,\omega \mathrm d\omega. 
\label{expansion_J_bar}
\end{equation}

The purpose of this paper is to derive a formal limit of this rescaled mean-field model when the parameter~$\varepsilon$ tends to~$0$. 
The first effect of this hydrodynamic scaling is that, up to order~$1$ in~$\varepsilon$, the equation becomes local.
Indeed, supposing that~$f^\varepsilon$ does not present any pathological behavior as~$\varepsilon\to0$, we get the following expansion:
\begin{equation}
\bar{J}^\varepsilon_{f^\varepsilon}(t,x)=J_{f^\varepsilon}(t,x)+O(\varepsilon^2),
\label{eq:expanJ}
\end{equation}
where the local flux~$J_f$ is defined by
\begin{equation}
J_f(x,t) = \int_{\mathbb{S}}  f(x,\omega,t) \,\omega \, \mathrm d\omega. 
\label{def_J}
\end{equation}
The proof of this expansion is elementary and omitted here (see e.g. Appendix~A.1 of~\cite{frouvelle2011continuum}).
We also define the density~$\rho_f$ associated to~$f$ by
\begin{equation}
\rho_f(x,t) = \int_{\mathbb{S}}  f(x,\omega,t) \, \mathrm d\omega. 
\label{def_rho2}
\end{equation}
Hence, Eq.~\eqref{KFP_eps} becomes, after dropping the~$O(\varepsilon^2)$ term: 
\begin{equation}
\varepsilon(\partial_t f^\varepsilon + \omega\cdot\nabla_xf^\varepsilon) = Q(f^\varepsilon) , 
\label{KFP_eps_red} 
\end{equation}
with
\begin{equation}
Q(f) = - \nabla_\omega \cdot ((\mbox{Id} - \omega \otimes \omega) J_f f) +  \Delta_\omega f.
\label{def_Q}
\end{equation}
This paper is concerned with the formal limit~$\varepsilon \to 0$ of this problem. 

We remark that the collision operator~$Q$ acts on the~$\omega$ variable only. 
The derivation of the macroscopic model relies on the properties of this operator. An obvious remark is that 
\begin{equation}
\int_{\omega \in {\mathbb S}} Q(f) \, d\omega = 0
\label{eq:int_Q}
\end{equation}
which expresses the local conservation of mass. The first step of the study consists in  characterizing the equilibria, i.e. the functions~$f$ such that~$Q(f)=0$. Indeed,  when~$\varepsilon\to0$,~$Q(f^\varepsilon)\to0$ and the limit~$f = \lim_{\varepsilon \to 0} f^\varepsilon$ belongs to the set of equilibria. This characterization is the purpose of the next subsection.

\subsection{Equilibria}
\label{subsec:equi}

For any unit vector~$\Omega \in \mathbb{S}$, and~$\kappa\geqslant0$, we define the so-called Von-Mises-Fisher distribution~\cite{watson1982distributions} with concentration parameter~$\kappa$ and orientation~$\Omega$ by \begin{equation}
 M_{\kappa\Omega}(\omega)=\frac{e^{\kappa\,\omega\cdot \Omega}}{\int_{\mathbb{S}}e^{\kappa\,\upsilon\cdot\Omega}\, \mathrm d \upsilon} \, .
\label{def_M}
\end{equation}
We note that the denominator depends only on~$\kappa$.
$M_{\kappa\Omega}$ is a probability density on the sphere, and we will denote by~$\langle\cdot\rangle_{M_{\kappa\Omega}}$ the average over this probability measure. For functions~$\gamma$ depending only on~$\omega\cdot\Omega$, the average~$\langle\gamma(\omega\cdot\Omega)\rangle_{M_{\kappa\Omega}}$ does not depend on~$\Omega$ and will be denoted by~$\langle\gamma(\cos\theta)\rangle_{M_\kappa}$. Using spherical coordinates, this average is given by: 
\begin{equation*}
\langle\gamma(\cos\theta)\rangle_{M_\kappa}=\frac{\int_0^\pi \gamma(\cos\theta)\, e^{\kappa \cos\theta}\sin^{n-2}\theta \,\mathrm d\theta}{\int_0^\pi e^{\kappa \cos\theta}\sin^{n-2}\theta \, \mathrm d\theta}.
\end{equation*}

The flux of the Von-Mises-Fisher distribution is given by
\begin{equation}
 J_{M_{\kappa\Omega}}=\langle\omega\rangle_{M_{\kappa\Omega}} = c(\kappa)\Omega,
\label{flux_M2}
\end{equation}
where the so-called order parameter~$c(\kappa)$ is such that~$0 \leqslant c(\kappa) \leqslant 1$ and is defined by
\begin{equation}
 c(\kappa)=\langle\cos\theta\rangle_{M_\kappa}=\frac{\int_0^\pi \cos\theta \, e^{\kappa \cos\theta}\sin^{n-2}\theta \, \mathrm d\theta}{\int_0^\pi e^{\kappa \cos\theta}\sin^{n-2}\theta \, \mathrm d\theta}.
\label{def_c}
\end{equation}
$c(\kappa)$ measures how the distribution~$M_{\kappa\Omega}$ is concentrated about~$\Omega$. When~$c(\kappa)=0$,~$M_{\kappa\Omega}$ is the uniform distribution~$M_{\kappa\Omega} = 1$, and when~$c(\kappa) \to 1$, we have~$M_{\kappa\Omega} \to \delta_\Omega(\omega)$.

We remark that the dependence of~$M_{\kappa\Omega}$ upon~$\kappa$ and~$\Omega$ only appears through the product~$\kappa\Omega$. In this way, we can consider~$M_{J}$ for any given vector~$J \in \mathbb{R}^n$. We also note that~$\nabla_\omega(M_{J})=(\mathrm{Id}-\omega\otimes\omega)J\, M_{J}$. Therefore
\begin{equation*}
Q(f) =\nabla_\omega \cdot \left[ M_{J_f} \nabla_\omega \left( \frac{f}{M_{J_f}} \right) \right].
\end{equation*}
Using Green's formula, we have
\begin{equation*}
\int_{\mathbb S} Q(f) \, \frac{g}{M_{J_f}} \, d\omega = - \int_{\mathbb S} \nabla_\omega \left( \frac{f}{M_{J_f}} \right) \cdot \nabla_\omega \left( \frac{g}{M_{J_f}} \right) \, M_{J_f} \, d\omega, 
\end{equation*}
and 
\begin{equation}
\int_{\mathbb S} Q(f) \, \frac{f}{M_{J_f}} \, d\omega = - \int_{\mathbb S} \left| \nabla_\omega \left( \frac{f}{M_{J_f}} \right) \right|^2 \, M_{J_f} \, d\omega \leqslant 0. 
\label{eq:intQf}
\end{equation}

\medskip
\begin{definition}
A function~$f(\omega)$ is said to be an equilibrium of~$Q$ if and only if~$Q(f)=0$.
\label{def:equi}
\end{definition}

Let~$f$ be an equilibrium. Using (\ref{eq:intQf}), we deduce that~$\frac{f}{M_{J_f}}$ is a constant. Therefore,~$f=\rho_f \, M_{J_f}$ is of the form~$\rho M_{\kappa\Omega}$ with~$\kappa\geqslant0$ and~$\Omega\in\mathbb{S}$ (we note that in the case~$|J_f|=0$, then~$\kappa=0$ and we can take any~$\Omega\in\mathbb{S}$ because~$f$  is then just the uniform distribution).
Using~\eqref{flux_M2}, we get~
$$\kappa\Omega=J_f=\rho J_{M_{\kappa\Omega}}=\rho c(\kappa)\Omega,$$ 
which leads to the following equation for~$\kappa$ (compatibility condition):
\begin{equation}
\rho c(\kappa)=\kappa.
\label{compatibility_condition}
\end{equation}
The study of this condition and the classification of the equilibria can be found in~\cite{frouvelle2011dynamics}.
The key point is to notice that the function~$\kappa\mapsto\frac{c(\kappa)}{\kappa}$ is decreasing and tends to~$\frac1n$ as~$\kappa\to0$. Therefore, there is no other solution than~$\kappa = 0$ if~$\rho \leqslant n$. By contrast, if~$\rho > n$, there is a unique strictly positive solution in addition to the trivial solution~$\kappa = 0$. This leads to the following proposition. 

\begin{proposition} (i) If~$\rho\leqslant n$,~$\kappa=0$ is the only solution to the compatibility relation (\ref{compatibility_condition}). The only equilibria are the isotropic ones~$f=\rho$, with arbitrary~$\rho\geqslant0$.

\medskip
\noindent
(ii)  If~$\rho>n$, the compatibility relation (\ref{compatibility_condition}) has exactly two roots:~$\kappa=0$ and a unique strictly positive root denoted by~$\kappa(\rho)$. The set of equilibria associated to the root~$\kappa=0$ consists of the isotropic equilibria ~$f=\rho$, with arbitrary~$\rho >n$. The set of equilibria associated to the root~$\kappa(\rho)$ consist of the Von Mises-Fischer distributions~$\rho M_{\kappa(\rho)\Omega}$ with arbitrary~$\rho >n$ and arbitrary~$\Omega \in {\mathbb S}$ and forms a manifold of dimension~$n$.  
\label{prop:equi}
\end{proposition}

The rate of convergence to the equilibria have been studied in ~\cite{frouvelle2011dynamics} in the spatially homogeneous setting. The results are recalled in the next section.

\subsection{Rates of convergence to equilibrium in the spatially homogeneous setting}
\label{subsection_convergence}

Denoting by~$g^\varepsilon = f^\varepsilon / \rho_{f^\varepsilon}$ the velocity probability distribution function, we can rewrite \eqref{KFP_eps_red} under the following form (omitting the superscripts~$\varepsilon$ for the sake of clarity and neglecting the~$O(\varepsilon^2)$ term):
\begin{equation*}
\varepsilon(\partial_t (\rho g) + \omega\cdot\nabla_x(\rho g)) = - (\rho)^2 \nabla_\omega \cdot ((\mbox{Id} - \omega \otimes \omega) J_{g} g) + \rho \Delta_\omega g .  
\end{equation*}
In the spatially homogeneous setting, we let~$\nabla_x(\rho g) = 0$ and get 
\begin{equation}
\varepsilon \partial_t (\rho g)  = - (\rho)^2 \nabla_\omega \cdot ((\mbox{Id} - \omega \otimes \omega) J_{g} g) + \rho \Delta_\omega g = Q(\rho g).  \label{eq:homo}
\end{equation}
Integrating this equation with respect to~$\omega$ and using (\ref{eq:int_Q}), we find that~$\partial_t \rho = 0$. Therefore,~$\rho$ is independent of~$t$ and can be cancelled out. The homogeneous equation (\ref{eq:homo}) therefore takes the form:
\begin{equation}
\varepsilon \partial_t g = - \rho \,  \nabla_\omega \cdot ((\mbox{Id} - \omega \otimes \omega) J_{g} g) +  \Delta_\omega g. \label{Doi_eq2} 
\end{equation}

We now remind the definitions of global and asymptotic rate. 

\begin{definition}
Let~${\mathcal X}$ be a Banach space with norm~$\| \cdot \|$ and let~$f(t)$:~${\mathbb R}_+ \to {\mathcal X}$ be a function of~$t$ with values in~${\mathcal X}$. We say that~$f(t)$ converges to~$f_\infty$ with global rate~$r$ if and only if there exists a constant~$C$ which only depends on~$\| f_0 \|$, such that 
\begin{equation} 
\| f(t) - f_\infty \| \leqslant C e^{-rt}. 
\label{eq:rate}
\end{equation}
We say that~$f(t)$ converges to~$f_\infty$ with asymptotic rate~$r$ if and only if there exists a constant~$C$ depending on~$f_0$ (but not only on~$\| f_0 \|$) such that (\ref{eq:rate}) holds. Finally, we say that~$f(t)$ converges to~$f_\infty$ with asymptotic algebraic rate~$\alpha$ if and only if there exists a constant~$C$ depending on~$f_0$
$$ \| f(t) - f_\infty \| \leqslant \frac{C}{t^\alpha}.~$$
\end{definition}

Now, concerning problem (\ref{Doi_eq2}), we can state the following theorem: 

\begin{theorem}
\label{thm_convergence}
{\bf \cite{frouvelle2011dynamics}} Suppose~$g_0$ is a probability measure, belonging to~$H^s(\mathbb{S})$.
There exists a unique weak solution~$g$ to (\ref{Doi_eq2}), with initial condition~$g(0)=g_0$.
Furthermore, this solution is a classical one, is positive for all time~$t>0$, and belongs to~$C^\infty((0,+\infty)\times\mathbb{S})$. 

\medskip
\noindent
(i) If~$J_{g_0}\neq0$, the large time behavior of the solution is given by one of the three cases below:

\smallskip
\noindent
- Case~$\rho < n$:~$g$ converges exponentially fast to the uniform distribution, with global  rate
\begin{equation} 
r(\rho) = \frac{(n-1)(n- \rho)}{n \varepsilon}, 
\label{eq:rate_rho<n}
\end{equation}
in any~$H^p$ norm.

\smallskip
\noindent
- Case~$\rho > n$:  There exists~$\Omega\in\mathbb{S}$ such that~$g$ converges exponentially fast to~$M_{\kappa(\rho)\Omega}$, with asymptotic rate greater than
\begin{equation*} 
r(\rho)=\frac{1}{\varepsilon} [\rho c(\kappa(\rho))^2 + n  - \rho] \Lambda_{\kappa(\rho)} > 0, 
\end{equation*}
in any~$H^p$ norm, where~$\Lambda_\kappa$ is the best constant for the following Poincar\'e inequality:
\begin{equation}
\langle|\nabla g|^2\rangle_{M_{\kappa\Omega}}\geqslant\Lambda_\kappa\langle(g-\langle g\rangle_{M_{\kappa\Omega}})^2\rangle_{M_{\kappa\Omega}},
\end{equation}
We have 
\begin{equation}
r(\rho) \sim \frac{1}{\varepsilon} 2(n-1)(\frac{\rho}{n} - 1), \quad \mbox{ when } \quad \rho \to n. 
\label{eq:asypt_rate}
\end{equation}

\smallskip
\noindent
- Case~$\rho = n$: then~$g$ converges to the uniform distribution in any~$H^p$ norm, with algebraic asymptotic rate~$1/2$. More precisely, we have:
$$ \|g-1\|_{H^p} \leq C \, \left( \frac{\varepsilon}{t}\right)^{1/2} . $$

\medskip
\noindent
(ii) If~$J_{g_0}=0$: Then, (\ref{Doi_eq2}) reduces to the heat equation on the sphere. So~$g$ converges to the uniform distribution, exponentially fast, with global rate~$r= \frac{2n}{\varepsilon}$ in any~$H^p$ norm. 
\label{thm:homogeneous_case}
\end{theorem}

\begin{remark}
That~$g_0$ is a probability measure implies that~$g_0 \in H^s(\mathbb{S})$ for all~$s<-\frac{n-1}{2}$. However, the theorem holds for all~$s$. So for~$s \geqslant -\frac{n-1}{2}$, that~$g_0 \in H^s(\mathbb{S})$ is not a mere consequence of being a probability measure and must added to the hypothesis. 
\label{rem:g0}
\end{remark}

Now, we comment the results of this theorem. First, in the supercritical case (when~$\rho>n$), the uniform distribution is an unstable equilibrium: for any perturbation~$g$ of the uniform distribution such that~$J_g \not = 0$, the associated solution converges to a given Von-Mises distribution, with a fixed concentration parameter~$\kappa(\rho)$ defined by the compatibility condition~\eqref{compatibility_condition}. Second, the rates of convergence to the equilibrium are exponential. In the supercritical case, these rates are only asymptotic ones, but  we can prove a uniform bound on these rates for~$\rho$ in any compact interval. A more precise study of the behavior of these rates is left to future work. 

Therefore, when~$\varepsilon$ is small, the function~$f^\varepsilon$ converges rapidly to a given equilibrium, provided that the rate satisfies~$r(\rho) \to \infty$ when~$\varepsilon \to 0$. In the case~$\rho<n$, thanks to (\ref{eq:rate_rho<n}), this condition is equivalent to saying that~$\varepsilon=o(n-\rho)$. In the case~$\rho>n$, thanks to (\ref{eq:asypt_rate}), the condition~$\varepsilon=o(n-\rho)$ implies that~$r(\rho) \to \infty$ when~$\varepsilon \to 0$ uniformly in any bounded~$\rho$ interval of the form~$[n,A]$ with~$A < \infty$. However, a uniform estimate from below of~$r(\rho)$ is lacking when~$\rho \to \infty$. But we can reasonably conjecture that away from a buffer region~$|\rho - n| = O(\varepsilon)$, the convergence to the equilibrium is exponentially fast.

\begin{figure}
\begin{center}
\begin{picture}(0,0)%
\includegraphics{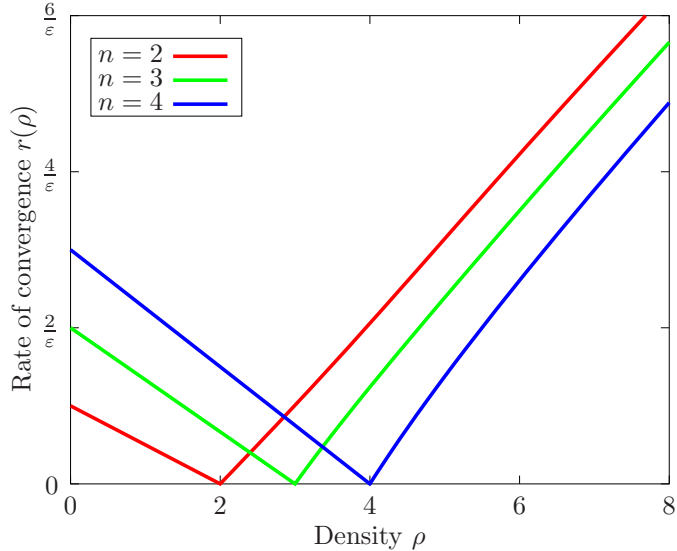}%
\end{picture}%
\setlength{\unitlength}{1973sp}%
\begingroup\makeatletter\ifx\SetFigFont\undefined%
\gdef\SetFigFont#1#2#3#4#5{%
  \reset@font\fontsize{#1}{#2pt}%
  \fontfamily{#3}\fontseries{#4}\fontshape{#5}%
  \selectfont}%
\fi\endgroup%
\begin{picture}(8303,6868)(1684,-7628)
\put(2298,-6894){\makebox(0,0)[rb]{\smash{{\SetFigFont{10}{12.0}{\familydefault}{\mddefault}{\updefault}0}}}}
\put(2448,-7144){\makebox(0,0)[b]{\smash{{\SetFigFont{10}{12.0}{\familydefault}{\mddefault}{\updefault}0}}}}
\put(4308,-7144){\makebox(0,0)[b]{\smash{{\SetFigFont{10}{12.0}{\familydefault}{\mddefault}{\updefault}2}}}}
\put(6168,-7144){\makebox(0,0)[b]{\smash{{\SetFigFont{10}{12.0}{\familydefault}{\mddefault}{\updefault}4}}}}
\put(8028,-7144){\makebox(0,0)[b]{\smash{{\SetFigFont{10}{12.0}{\familydefault}{\mddefault}{\updefault}6}}}}
\put(9888,-7144){\makebox(0,0)[b]{\smash{{\SetFigFont{10}{12.0}{\familydefault}{\mddefault}{\updefault}8}}}}
\put(3601,-2086){\makebox(0,0)[rb]{\smash{{\SetFigFont{10}{12.0}{\familydefault}{\mddefault}{\updefault}$n=4$}}}}
\put(3601,-1786){\makebox(0,0)[rb]{\smash{{\SetFigFont{10}{12.0}{\familydefault}{\mddefault}{\updefault}$n=3$}}}}
\put(3601,-1486){\makebox(0,0)[rb]{\smash{{\SetFigFont{10}{12.0}{\familydefault}{\mddefault}{\updefault}$n=2$}}}}
\put(1951,-3811){\rotatebox{90.0}{\makebox(0,0)[b]{\smash{{\SetFigFont{10}{12.0}{\familydefault}{\mddefault}{\updefault}Rate of convergence $r(\rho)$}}}}}
\put(2298,-1027){\makebox(0,0)[rb]{\smash{{\SetFigFont{10}{12.0}{\familydefault}{\mddefault}{\updefault}$\frac6\varepsilon$}}}}
\put(2298,-2983){\makebox(0,0)[rb]{\smash{{\SetFigFont{10}{12.0}{\familydefault}{\mddefault}{\updefault}$\frac4\varepsilon$}}}}
\put(2298,-4938){\makebox(0,0)[rb]{\smash{{\SetFigFont{10}{12.0}{\familydefault}{\mddefault}{\updefault}$\frac2\varepsilon$}}}}
\put(6168,-7519){\makebox(0,0)[b]{\smash{{\SetFigFont{10}{12.0}{\familydefault}{\mddefault}{\updefault}Density $\rho$}}}}
\end{picture}%
\caption{Rates of convergence to equilibria in dimensions~$2$,~$3$, and~$4$, as functions of the density~$\rho$.}
\label{fig_rates}
\end{center}
\end{figure}

Some elements towards a uniform estimate of the rate~$r(\rho)$ are provided in Appendix 1. 
Furthermore, in Appendix 2, we compute~$\Lambda_\kappa$ and then~$r(\rho)$ numerically. The results are depicted in Fig.~\ref{fig_rates} for dimensions~$2$,~$3$, and~$4$. 
We observe that for~$\rho>n$,~$r(\rho)$ grows linearly with~$\rho$, which supports our conjecture. 

Therefore, in the general space-inhomogeneous case, we will assume that the formal limit  of~$f^\varepsilon$ as~$\varepsilon\to 0$ is given by a function~$f(x,\omega,t)$ which has a different velocity profile according to the position of the local density~$\rho(x,t)$ with respect to the threshold value~$n$. For this purpose, we define the disordered region~$\mathcal{R}_d$ and the ordered region~$\mathcal{R}_o$ as 
\begin{eqnarray} 
& &  \mathcal{R}_d = \{ (x,t) \, \, | \, \, n-\rho^\varepsilon(x,t) \gg \varepsilon, \quad \mbox{ as } \varepsilon \to 0 \, \}, \label{eq:Rd}\\
& &  \mathcal{R}_o = \{ (x,t) \, \, | \, \, \rho^\varepsilon(x,t)-n \gg \varepsilon, \quad \mbox{ as } \varepsilon \to 0 \,  \}. \label{eq:Ro}
\end{eqnarray} 
We assume that as~$\varepsilon \to 0$ we have
\begin{eqnarray} 
& &  f^\varepsilon (x,\omega,t) \to \rho(x,t), \quad \forall (x,t) \in \mathcal{R}_d, \label{eq:equi_d}\\
& & f^\varepsilon (x,\omega,t) \to \rho(x,t) \, M_{\kappa(\rho)\Omega(x,t)},  \quad \forall (x,t) \in \mathcal{R}_o,   \label{eq:equi_o}
\end{eqnarray}
and that the convergence is as smooth as needed.  

The goal is now to derive evolution equations for ~$\rho(x,t)$ and ~$\Omega(x,t)$. This is the subject of the following two sections. We already note that, integrating \eqref{KFP_eps_red} with respect to~$\omega$ and using (\ref{eq:int_Q}), we get the mass conservation equation
\begin{equation}
\partial_t \rho^\varepsilon+\nabla_x\cdot(J_{f^\varepsilon}) = 0.
\label{mass_conservation_eps_red}
\end{equation}

\setcounter{equation}{0}
\section{Diffusion model in the disordered region}
\label{section-disorder}

We derive the macroscopic model in the disordered region~$\mathcal{R}_d\subset\mathbb{R}^n$, using (\ref{eq:equi_d}). With~\eqref{mass_conservation_eps_red} and the fact that ~$J_{f^\varepsilon}\to J_{f}=0$, the mass conservation equation reduces to 
\begin{equation*}
\partial_t\rho=0.
\end{equation*}
To obtain more precise information, we look for the next order in~$\varepsilon$, using a Chapman-Enskog method, similarly to the case of rarefied gas dynamics (see~\cite{degond2004macroscopic} for a review). We prove the following theorem:

\begin{theorem}
When~$\varepsilon$ tends to zero, the (formal) first order approximation to the solution of the rescaled mean-field system~(\ref{KFP_eps_red}), (\ref{def_Q}) in the disordered region~$\mathcal{R}_d$ defined by (\ref{eq:Rd}) is given by~
\begin{equation}
f^\varepsilon(x,\omega,t)= \rho^\varepsilon(x,t)-\varepsilon\,\frac{n\,\omega\cdot\nabla_x\rho^\varepsilon(x,t)}{(n-1)(n-\rho^\varepsilon(x,t))},
\label{eq:diff_correction}
\end{equation}
where the density~$\rho^\varepsilon$ satisfies the following diffusion equation
\begin{equation}
\partial_t \rho^\varepsilon = \frac{\varepsilon}{n-1}\nabla_x\cdot\left(\frac1{n-\rho^\varepsilon} \, \nabla_x\rho^\varepsilon\right).
\label{diffusion_disorder}
\end{equation}
\label{thm:limit_disorder}
\end{theorem}

\medskip
\noindent
{\bf Proof.} We let~$\rho^\varepsilon = \rho_{f^\varepsilon}$ and write~$f^\varepsilon=\rho^\varepsilon(x,t)+\varepsilon f_1^\varepsilon(x,\omega,t)$ with~$\int_\mathbb{S}f_1^\varepsilon\mathrm d \omega=0$. Inserting this Ansatz into (\ref{def_J}), we get 
\begin{equation*}
J^\varepsilon_{f^\varepsilon}=\varepsilon J_{f_1^\varepsilon}(t,x),
\end{equation*}
and the model (\ref{KFP_eps_red}), (\ref{def_Q}) becomes:
\begin{equation}
\begin{split}
\partial_t\rho^\varepsilon+\omega\cdot\nabla_x\rho^\varepsilon+\varepsilon(\partial_t+\omega\cdot\nabla_x)f_1^\varepsilon&=-\nabla_\omega((\mathrm{Id}-\omega\otimes\omega)J_{f_1^\varepsilon}\rho^\varepsilon) + \Delta_\omega f_1^\varepsilon \\
&\hspace{5ex}- \varepsilon \nabla_\omega((\mathrm{Id}-\omega\otimes\omega)J_{f_1^\varepsilon}\rho^\varepsilon).
\end{split}
\label{KFP_disorder}
\end{equation}
Additionally, (\ref{mass_conservation_eps_red}) gives:
\begin{equation}
\partial_t \rho^\varepsilon+\varepsilon\nabla_x\cdot(J_{f_1^\varepsilon}) = 0.
\label{mass_conservation_disorder}
\end{equation}
In particular~$\partial_t\rho^\varepsilon = O(\varepsilon)$. We need to compute~$f_1^\varepsilon$ to find the expression of the current. But, with this aim, we may retain only the terms of order~$0$ in~\eqref{KFP_disorder}. Since
$$ \nabla_\omega((\mathrm{Id}-\omega\otimes\omega)A)=-(n-1)A\cdot\omega,~$$ 
for any constant vector~$A\in\mathbb{R}^n$, the equation for~$f_1^\varepsilon$ reads: 
\begin{equation*}
\Delta_\omega f^\varepsilon_1=(\nabla_x\rho^\varepsilon-(n-1)\rho^\varepsilon J_{f^\varepsilon_1})\cdot\omega+O(\varepsilon).
\end{equation*}
This equation can be easily solved, since the right-hand side is a spherical harmonic of degree~$1$ (i.e. is of the form~$A\cdot\omega$; we recall that ~$\Delta_\omega(A\cdot\omega)=-(n-1)A\cdot\omega$ and that~$A \cdot \omega$ is of zero mean). Then:
\begin{equation*}
f^\varepsilon_1=-\frac1{n-1}(\nabla_x\rho^\varepsilon-(n-1)\rho^\varepsilon J_{f^\varepsilon_1})\cdot\omega+O(\varepsilon).
\end{equation*}
We immediately deduce, using that ~$\int_\mathbb{S}\omega\otimes\omega\mathrm d \omega=\frac1n\mathrm{Id}$:  
\begin{equation*}
J_{f^\varepsilon_1}=\frac{-1}{n(n-1)}(\nabla_x\rho^\varepsilon-(n-1)\rho^\varepsilon J_{f^\varepsilon_1})+O(\varepsilon),
\end{equation*}
which implies that 
\begin{equation*}
J_{f^\varepsilon_1}=\frac{-1}{(n-1)(n-\rho^\varepsilon)}(\nabla_x\rho^\varepsilon+O(\varepsilon)).
\end{equation*}
Inserting this equation into (\ref{mass_conservation_disorder}) leads to the diffusion model (\ref{diffusion_disorder}) and ends the proof. \endproof

\begin{remark}
The expression of~$f^\varepsilon_1$, which is given by the~$O(\varepsilon)$ term of (\ref{eq:diff_correction}) confirms that the approximation is only valid when~$n-\rho^\varepsilon \gg \varepsilon$.
The diffusion coefficient is only positive in the disordered region and it blows up as~$\rho^\varepsilon$ tends to~$n$, showing that the Chapman-Enskog expansion loses its validity. 
\end{remark}

\setcounter{equation}{0}
\section{Hydrodynamic model in the ordered region}
\label{section-order}

\subsection{Derivation of the model}
\label{subsec_deriv}

We now turn to the ordered region~$\mathcal{R}_o\subset\mathbb{R}^n$ defined by (\ref{eq:Ro}). The purpose of this section is to give a formal proof of the following:

\begin{theorem}
\label{prop_hydrodynamic}
When~$\varepsilon$ tends to zero, the (formal) limit to the solution~$f^\varepsilon(x,\omega,t)$ of the rescaled mean-field system~(\ref{KFP_eps_red}), (\ref{def_Q}), in the ordered region~$\mathcal{R}_o\subset\mathbb{R}^n$ defined by (\ref{eq:Ro}), is given by
\begin{equation}
f(x,\omega,t)= \rho(x,t) \, M_{\kappa(\rho(x,t))\Omega(x,t)}(\omega),
\label{eq:equi_stable}
\end{equation}
where the Von-Mises-Fischer distribution~$M_{\kappa\Omega}$ is defined at~\eqref{def_M}, and the parameter~$\kappa$ is the unique positive solution to the compatibility condition~\eqref{compatibility_condition}.
Moreover, the density~$\rho>n$ and the orientation~$\Omega\in\mathbb{S}$ satisfy the following system of first order partial differential equations:
\begin{gather}
\partial_t \rho + \nabla_x\cdot(\rho c \Omega)= 0,\label{mass_conservation_order}\\
\label{orientation_evolution_order}
\rho(\partial_t\Omega+\widetilde c (\Omega\cdot\nabla_x)\Omega)+\lambda(\mathrm{Id}-\Omega\otimes\Omega) \nabla_x \rho =0,
\end{gather}
where the coefficient~$c=c(\kappa(\rho))$ is defined at~\eqref{def_c}, the coefficient~$\widetilde c = \widetilde c(\kappa(\rho))$ will be defined later on at~\eqref{def_ctild}, and the parameter~$\lambda = \lambda(\rho)$ is given by
\begin{equation}
\lambda=\frac{\rho-n-\kappa\widetilde c}{\kappa(\rho-n-\kappa c)}.\label{def_lambda2}
\end{equation}
\end{theorem}

\medskip
\noindent
{\bf Proof:} From now on, we will drop the dependence on~$\rho$ in the coefficients when no confusion is possible. With (\ref{eq:equi_o}),~$f^\varepsilon \to f$, where~$f$ is the stable local equilibrium (\ref{eq:equi_stable}). We now derive the evolution equations (\ref{mass_conservation_order}), (\ref{orientation_evolution_order}) for~$\rho$ and~$\Omega$. 

We recall that the concentration parameter~$\kappa$  satisfies the compatibility condition~\eqref{compatibility_condition} where the order parameter~$c$ is defined by~\eqref{def_c} and that we have~$J_f=\rho c\Omega$. Therefore, eq.~\eqref{mass_conservation_eps_red} in the limit~$\varepsilon\to0$, reads
\begin{equation*}
\partial_t \rho + \nabla_x\cdot(\rho c \Omega)= 0.
\end{equation*}

To compute the evolution equation for~$\Omega$, the method proposed originally in~\cite{degond2008continuum} consists in introducing the notion of generalized collisional invariant (GCI). This method has been then applied to~\cite{degond2010macroscopic, frouvelle2011continuum}.
The first step is the definition and determination of the GCI's.
We define the linear operator~$L_{\kappa\Omega}$ associated to a concentration parameter~$\kappa$ and a direction~$\Omega$ as follows:
\begin{equation*}
L_{\kappa\Omega}(f) = \Delta_\omega f - \kappa\nabla_\omega\cdot((\mathrm{Id}-\omega\otimes\omega)\Omega f)=\nabla_\omega \cdot \left[ M_{\kappa\Omega} \nabla_\omega \left( \frac{f}{M_{\kappa\Omega}} \right) \right],
\end{equation*}
so that~$Q(f)=L_{J_f}(f)$. We define the set~${\mathcal C}_{\kappa\Omega}$ of GCI's associated to~$\kappa\in\mathbb{R}$ and~$\Omega\in\mathbb{S}$ by:
\begin{equation*} 
{\mathcal C}_{\kappa\Omega}=\left\{\psi|\int_{\omega \in \mathbb{S}} L_{\kappa\Omega}(f) \, \psi \, \mathrm d\omega = 0 , \, \forall f \text{ such that } \, (\mathrm{Id}-\Omega\otimes\Omega)J_f=0 \right\}.
\end{equation*}
Hence, if~$\psi$ is a GCI associated to~$\kappa$ and~$\Omega$, we have:
$$ \int_{\omega \in \mathbb{S}} Q(f) \, \psi \, \mathrm d\omega = 0, \quad  \forall f \mbox{ such that } J_f=\kappa\Omega.~$$

The determination of~${\mathcal C}_{\kappa\Omega}$ closely follows~\cite{frouvelle2011continuum}. 
We define the space
\begin{equation}
V = \{ g \, | \,(n-2)(\sin\theta)^{\frac n2-2} g \in L^2(0,\pi), \, (\sin\theta)^{\frac n2-1}g \in H^1_0(0,\pi) \},
\label{def_V2}
\end{equation}
and we denote by~$g_\kappa$ the unique solution in~$V$ of the elliptic problem~
\begin{equation}
\widetilde L_\kappa^*g(\theta)=\sin\theta, 
\label{def_GCI_elliptic}
\end{equation}
where
\begin{equation}
\widetilde L_\kappa^*g(\theta)=-(\sin \theta)^{2-n} e^{-\kappa\cos\theta}\tfrac{\mathrm d}{\mathrm d\theta} \big( (\sin \theta)^{n-2} e^{\kappa\cos\theta} \frac{dg}{d \theta} (\theta) \big)+\tfrac{n-2}{\sin^2\theta}\,g(\theta).
\label{def_Ltild}
\end{equation}
Then defining~$h_\kappa$ by~$g_\kappa(\theta)=h_\kappa(\cos\theta) \, \sin\theta$, we get 
$$ {\mathcal C}_{\kappa\Omega} = \{ h_\kappa(\omega\cdot\Omega)A\cdot\omega+C \quad | \quad
C\in\mathbb{R}, \quad A\in\mathbb{R}^n, \mbox{ with }  A\cdot \Omega = 0 \, \} .$$
${\mathcal C}_{\kappa\Omega}$ is a vector space of dimension~$n$, since~$A$ is a vector with~$n-1$ independent components. 

The next step consists in multiplying~\eqref{KFP_eps_red} by a GCI associated to~$\kappa^\varepsilon$ and~$\Omega^\varepsilon$ such that~$J_{f^\varepsilon}=\kappa^\varepsilon\Omega^\varepsilon$, and to integrate it with respect to~$\omega$. For any vector~$A\in\mathbb{R}^n$, with~$A\cdot\Omega^\varepsilon=0$, we get
\begin{equation*}
\int_{\omega \in \mathbb{S}}Q(f^\varepsilon)h_{\kappa^\varepsilon}(\omega\cdot\Omega^\varepsilon) \,A\cdot\omega\, \mathrm d\omega=0.
\end{equation*}
So, the vector~
$$ X^\varepsilon=\frac1{\varepsilon}\int_{\omega \in \mathbb{S}}Q(f^\varepsilon)h_{\kappa^\varepsilon}(\omega\cdot\Omega^\varepsilon)\, \omega\, \mathrm d\omega ,~$$ 
is parallel to~$\Omega^\varepsilon$, or equivalently~$(\mathrm{Id} - \Omega^\varepsilon \otimes \Omega^\varepsilon)\, X^\varepsilon=0$.
Using~\eqref{KFP_eps_red}, we get:
\begin{equation*} 
X^\varepsilon = \int_{\omega \in \mathbb{S}} ( \partial_t f^\varepsilon + \omega \cdot \nabla_x f^\varepsilon)\, h_{\kappa^\varepsilon}(\omega \cdot \Omega^\varepsilon) \, \omega \, \mathrm d\omega.
\end{equation*}
In the limit~$\varepsilon\to0$, we get
\begin{equation} 
(\mathrm{Id} - \Omega \otimes \Omega)\, X=0, 
\label{eq:(I-OO)X}
\end{equation}
where 
\begin{equation*} 
X = \int_{\omega \in \mathbb{S}} ( \partial_t ( \rho M_{\kappa\Omega}) + \omega \cdot \nabla_x (\rho M_{\kappa\Omega}))\, h_\kappa(\omega \cdot \Omega) \, \omega \, \mathrm d\omega\, .
\end{equation*}
Finally it has been proved in~\cite{frouvelle2011continuum} that (\ref{eq:(I-OO)X}) is equivalent to (\ref{orientation_evolution_order}) with
\begin{align}
\widetilde c&=\langle \cos\theta \rangle_{\widetilde M_\kappa}=
 \frac{\int_0^\pi \cos \theta h_\kappa(\cos\theta) e^{\kappa\cos\theta} \, 
\sin^{n} \theta \, \mathrm d\theta}{\int_0^\pi h_\kappa(\cos\theta)e^{\kappa\cos\theta} \, \sin^{n} \theta \, \mathrm d\theta}\, ,\label{def_ctild} \\
\lambda&=\frac1{\kappa}+\frac{\rho}{\kappa}\,\frac{\mathrm d \kappa}{\mathrm d \rho} \,(\widetilde c -c)\, .
\end{align}

We can now compute a simpler expression of~$\lambda$. We differentiate the compatibility condition (\ref{compatibility_condition}) with respect to~$\kappa$, and we get
$$c\frac{\mathrm d \rho}{\mathrm d \kappa}+\rho\frac{\mathrm d c}{\mathrm d \kappa}=1.$$ 
We have
\begin{align*}
\frac{\mathrm d c}{\mathrm d \kappa}&=\frac{\mathrm d}{\mathrm d \kappa}\left(\frac{\int_0^\pi \cos\theta \, e^{\kappa \cos\theta}\sin^{n-2}\theta \, \mathrm d\theta}{\int_0^\pi e^{\kappa \cos\theta}\sin^{n-2}\theta \, \mathrm d\theta}\right)\\
&=\frac{\int_0^\pi \cos^2\theta \, e^{\kappa \cos\theta}\sin^{n-2}\theta \, \mathrm d\theta}{\int_0^\pi e^{\kappa \cos\theta}\sin^{n-2}\theta \, \mathrm d\theta}-\left(\frac{\int_0^\pi \cos\theta \, e^{\kappa \cos\theta}\sin^{n-2}\theta \, \mathrm d\theta}{\int_0^\pi e^{\kappa \cos\theta}\sin^{n-2}\theta \, \mathrm d\theta}\right)^2\\
&=1-\frac{\int_0^\pi \sin^2\theta \, e^{\kappa \cos\theta}\sin^{n-2}\theta \, \mathrm d\theta}{\int_0^\pi e^{\kappa \cos\theta}\sin^{n-2}\theta \, \mathrm d\theta}-c^2\\
&=1-(n-1)\frac{c}{\kappa}-c^2.
\end{align*}
Therefore we get
\begin{equation}
c\,\frac{\mathrm d \rho}{\mathrm d \kappa}=\frac{\kappa}{\rho}\,\frac{\mathrm d \rho}{\mathrm d \kappa}=
1 - \rho \frac{dc}{d\kappa} = 
1-\rho\,(1-(n-1)\frac{c}{\kappa}-c^2)=n-\rho+\kappa c,
\label{eq:ugly}
\end{equation}
and finally
\begin{equation*}
\lambda=\frac1{\kappa}+\frac{\widetilde c-c}{n-\rho+\kappa c}=\frac{n-\rho+\kappa\widetilde c}{\kappa(n-\rho+\kappa c)},
\end{equation*}
which ends the proof of Theorem~\ref{prop_hydrodynamic}. \endproof

The next part is devoted to the study of the properties of the model~\eqref{mass_conservation_order}-\eqref{orientation_evolution_order} in the ordered region.

\subsection{Hyperbolicity of the hydrodynamic model in the ordered region}
\label{subsec:properties}

We first investigate the hyperbolicity of the hydrodynamic model~\eqref{mass_conservation_order}-\eqref{orientation_evolution_order}. We recall some definitions. Let 
\begin{equation}
\partial_t U + \sum_{i=1}^n A_i(U) \partial_{x_i} U = 0 ,
\label{eq:first_order_system}
\end{equation}
be a first order system where~$x \in {\mathbb R}^n$,~$t \geqslant0$,~$U=(U_1, \ldots U_m)$ is a~$m$-dimensional vector and~$(A_i(U))_{i=1, \ldots,n}$ are~$n$~$m \times m$-dimensional matrices. 
Let~$U_0  \in {\mathbb R}^m$. The constant and uniform state~$U(x,t) = U_0$ is a particular solution of (\ref{eq:first_order_system}). The linearization of (\ref{eq:first_order_system}) about this constant and uniform state leads to the following linearized system: 
\begin{equation}
\partial_t u + \sum_{i=1}^n A_i(U_0) \partial_{x_i} u = 0 .
\label{eq:first_order_system_linearized}
\end{equation}
We look for solutions of (\ref{eq:first_order_system_linearized}) in the form of plane waves~$u(x,t) = \bar u \, e^{i(k \cdot x - \omega t)}$, with~$k \in {\mathbb R}^n$ and~$\omega \in {\mathbb C}$. Such solutions exist if and only if~$\omega/|k|$ is an eigenvalue of the matrix~$A(k/|k|)$ and~$\bar u$ is the related eigenvector, where for a direction~$\xi \in {\mathbb S}$, the matrix~$A(\xi)$ is defined by
\begin{equation}
A(\xi) = \sum_{i=1}^n A_i(U) \xi_i.  
\label{eq:A(xi)}
\end{equation}
The problem (\ref{eq:first_order_system}) is said to be hyperbolic about~$U_0$, if only purely propagative plane waves with real~$\omega$ can exist or equivalently, if~$A(\xi)$ has real eigenvalues for any~$\xi$. We also must rule out polynomially increasing in time solutions which could exist if the matrix would not be diagonalizable. This leads to the following definitions:  

\begin{definition}
(i) Let~$U_0  \in {\mathbb R}^m$. System (\ref{eq:first_order_system}) is hyperbolic about~$U_0$ if and only if for all directions~$\xi \in {\mathbb S}$, the matrix~$A(\xi)$ is diagonalizable with real eigenvalues. \\

\noindent
(ii) System (\ref{eq:first_order_system}) is hyperbolic, if and only if it is hyperbolic about any state~$U_0$ in the domain of definition of the matrices~$A_i(U)$. 

\label{def_hyperbolicity}
\end{definition}

The linearization of system~\eqref{mass_conservation_order}-\eqref{orientation_evolution_order} about a stationary uniform state~$(\rho_0,\Omega_0)$ is obtained by inserting the following expansion
\begin{eqnarray}
& & \hspace{-1cm} \rho = \rho_0 + \delta r + o(\delta), \label{eq:expan_rho} \\
& & \hspace{-1cm} \Omega = \Omega_0 + \delta W + o(\delta), \label{eq:expan_Omega} 
\end{eqnarray}
with~$\delta \ll 1$ a small parameter and~$r=r(x,t)$,~$W=W(x,t)$, the first order perturbations of~$\rho$ and~$\Omega$. Given that~$|\Omega| = |\Omega_0| = 1$, we have~$ W \cdot \Omega_0 = 0$. 
Inserting (\ref{eq:expan_rho}), (\ref{eq:expan_Omega}) into~\eqref{mass_conservation_order}-\eqref{orientation_evolution_order} leads to the following linearized system: 
\begin{eqnarray}
& & \hspace{-1cm} 
\partial_t r + \gamma_0 (\Omega_0 \cdot \nabla_x) r + \rho_0 c_0 (\nabla_x \cdot W) = 0, \label{mass_conservation_order_linearized}\\
& & \hspace{-1cm} \partial_t W + \widetilde c_0  (\Omega_0\cdot\nabla_x)W + \frac{\lambda_0}{\rho_0} (\mathrm{Id}-\Omega_0\otimes\Omega_0) \nabla_x r=0, \label{orientation_evolution_order_linearized} \\
& & \hspace{-1cm} 
W \cdot \Omega_0 = 0 , \label{eq:cnd_linearized}
\end{eqnarray}
with 
$$ \gamma(\rho) = c + \rho \frac{dc}{d\rho},~$$
and~$\gamma_0 = \gamma(\rho_0)$,~$c_0 = c(\rho_0)$,~$\tilde c_0 = \tilde c(\rho_0)$ and~$\lambda_0 = \lambda(\rho_0)$. 

Next, we show that system (\ref{mass_conservation_order_linearized})-(\ref{eq:cnd_linearized}) is invariant under rotations. This will allow us to choose one arbitrary direction~$\xi$ in the definition (\ref{eq:A(xi)}) instead of checking all possible directions. For this purpose, let~$R$ be a rotation matrix of~${\mathbb R}^n$, i.e.~$R$ is a~$n \times n$ matrix such that~$R^T = R^{-1}$, where the exponent~$T$ denotes transposition. We introduce the change of variables~$x = R x'$ and define new unknowns 
\begin{eqnarray*}
& & \hspace{-1cm} 
r(x) = r'(x'), \quad W(x) = R W'(x'), \quad \Omega_0 = R \Omega_0'. 
\end{eqnarray*}
We note the following identities
\begin{eqnarray*}
& & \hspace{-1cm} 
\Omega_0' \cdot W'(x') = \Omega_0 \cdot W(x) = 0 , \\
& & \hspace{-1cm} 
\nabla_x r(x) = R \, \nabla_{x'} r'(x') , \\
& & \hspace{-1cm} 
\nabla_x W(x) = R \, \nabla_{x'} W'(x')\,  R^T , \\
& & \hspace{-1cm} 
(\nabla_x \cdot W)(x) = (\nabla_{x'} \cdot W')(x'), \\
& & \hspace{-1cm} 
(\Omega_0 \cdot \nabla_x) W(x) = (\nabla_x W(x))^T \Omega_0 = R \, (\nabla_{x'} W(x'))^T \Omega_0' =  R \,  (\Omega_0' \cdot \nabla_{x'}) W'(x') , \\
& & \hspace{-1cm} 
(\Omega_0 \cdot \nabla_x) r(x) = (\Omega_0' \cdot \nabla_{x'}) r'(x') . 
\end{eqnarray*}
With these identities, it is easy to show that~$(r',W')$ satisfies system (\ref{mass_conservation_order_linearized})-(\ref{eq:cnd_linearized}) with~$\Omega_0$ replaced by~$\Omega'_0$. 

The rotational invariance of (\ref{mass_conservation_order_linearized})-(\ref{eq:cnd_linearized}) shows that, in order to check the hyperbolicity, it is enough to choose any particular direction~$\xi$. Let us call this arbitrary direction~$z$, with unit vector in this direction denoted by~$e_z$. To check the hyperbolicity of waves propagating in the~$z$ direction it is sufficient to look at the system where all unknowns only depend only on the space coordinate~$z$ and on the time~$t$. Denoting by~$\theta$ the angle between the~$z$ direction and~$\Omega$, we can write: 
$$\Omega=\cos\theta\,e_z+\sin\theta\,v, \quad \theta \in [0,\pi], \quad  v\in\mathbb{S}_{n-2},~$$ 
where~$\mathbb{S}_{n-2}$ is the sphere of dimension~$n-2$ collecting all unit vectors orthogonal to~$e_z$. With these hypotheses, system~\eqref{mass_conservation_order}-\eqref{orientation_evolution_order} is written.
\begin{gather} 
\partial_t \rho + \, \partial_z (\rho c(\rho) \cos \theta) = 0.
\label{rho_eq_z2} \\
\rho[\partial_t (\cos\theta) + \tilde c(\rho) \cos \theta \, \partial_z (\cos\theta)] + \lambda \, \sin^2 \theta \, \partial_z \rho = 0.
\label{theta_eq_z2} \\
\partial_t v + \tilde c(\rho) \cos \theta \, \partial_z v = 0, \text{ with }|v|=1 \text{ and } e_z\cdot v=0.
\label{v_eq_z2}
\end{gather}
In the special case of dimension~$n=2$, the system reduces to~\eqref{rho_eq_z2}-\eqref{theta_eq_z2}, with~$\theta\in(-\pi,\pi)$ and~$\Omega=\cos\theta\,e_z+\sin\theta\,v_0$, where~$v_0$ is one of the two unit vectors orthogonal to~$e_z$.

\begin{figure}
\begin{center}
\begin{picture}(0,0)%
\includegraphics{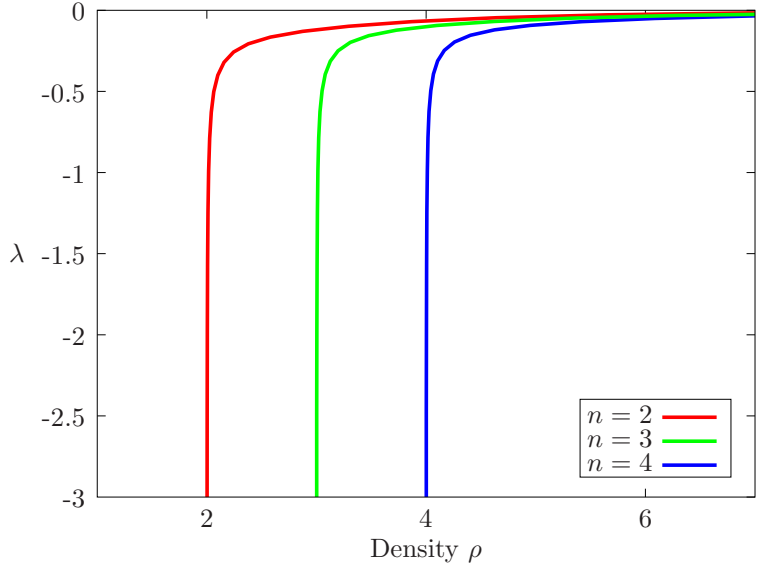}%
\end{picture}%
\setlength{\unitlength}{1973sp}%
\begingroup\makeatletter\ifx\SetFigFont\undefined%
\gdef\SetFigFont#1#2#3#4#5{%
  \reset@font\fontsize{#1}{#2pt}%
  \fontfamily{#3}\fontseries{#4}\fontshape{#5}%
  \selectfont}%
\fi\endgroup%
\begin{picture}(9133,7073)(1261,-7620)
\put(9076,-6436){\makebox(0,0)[rb]{\smash{{\SetFigFont{10}{12.0}{\familydefault}{\mddefault}{\updefault}$n=4$}}}}
\put(9076,-6136){\makebox(0,0)[rb]{\smash{{\SetFigFont{10}{12.0}{\familydefault}{\mddefault}{\updefault}$n=3$}}}}
\put(9076,-5836){\makebox(0,0)[rb]{\smash{{\SetFigFont{10}{12.0}{\familydefault}{\mddefault}{\updefault}$n=2$}}}}
\put(1276,-3811){\makebox(0,0)[rb]{\smash{{\SetFigFont{10}{12.0}{\familydefault}{\mddefault}{\updefault}$\lambda$}}}}
\put(6263,-7511){\makebox(0,0)[b]{\smash{{\SetFigFont{10}{12.0}{\familydefault}{\mddefault}{\updefault}Density $\rho$}}}}
\put(6263,-7136){\makebox(0,0)[b]{\smash{{\SetFigFont{10}{12.0}{\familydefault}{\mddefault}{\updefault}4}}}}
\put(8988,-7136){\makebox(0,0)[b]{\smash{{\SetFigFont{10}{12.0}{\familydefault}{\mddefault}{\updefault}6}}}}
\put(3538,-7136){\makebox(0,0)[b]{\smash{{\SetFigFont{10}{12.0}{\familydefault}{\mddefault}{\updefault}2}}}}
\put(2026,-787){\makebox(0,0)[rb]{\smash{{\SetFigFont{10}{12.0}{\familydefault}{\mddefault}{\updefault}0}}}}
\put(2026,-1803){\makebox(0,0)[rb]{\smash{{\SetFigFont{10}{12.0}{\familydefault}{\mddefault}{\updefault}-0.5}}}}
\put(2026,-2820){\makebox(0,0)[rb]{\smash{{\SetFigFont{10}{12.0}{\familydefault}{\mddefault}{\updefault}-1}}}}
\put(2026,-3836){\makebox(0,0)[rb]{\smash{{\SetFigFont{10}{12.0}{\familydefault}{\mddefault}{\updefault}-1.5}}}}
\put(2026,-4853){\makebox(0,0)[rb]{\smash{{\SetFigFont{10}{12.0}{\familydefault}{\mddefault}{\updefault}-2}}}}
\put(2026,-5869){\makebox(0,0)[rb]{\smash{{\SetFigFont{10}{12.0}{\familydefault}{\mddefault}{\updefault}-2.5}}}}
\put(2026,-6886){\makebox(0,0)[rb]{\smash{{\SetFigFont{10}{12.0}{\familydefault}{\mddefault}{\updefault}-3}}}}
\end{picture}%
\caption{Coefficient~$\lambda$ in dimensions~$2$,~$3$, and~$4$.}
\label{fig_lambda}
\end{center}
\end{figure}

The hyperbolicity of this system depends on the sign of~$\lambda$. Proposition \ref{prop:hyperbolic_expansion} below shows that~$\lambda <0$ in the two limits~$\rho \to n$ and~$\rho \to \infty$. Additionally, the numerical computation of~$\lambda$, displayed in Fig.~\ref{fig_lambda}, provides evidence that~$\lambda<0$ for all values of~$\rho$, at least in dimensions~$n=2$,~$3$, and~$4$. Therefore, we assume that 
\begin{equation} 
\lambda <0.
\label{eq:lambda<0}
\end{equation}

We first check the local hyperbolicity criterion: 

\begin{proposition}
We assume (\ref{eq:lambda<0}). Then, system (\ref{rho_eq_z2})-(\ref{v_eq_z2}) is hyperbolic about~$(\rho, \theta, v)$ if and only if 
\begin{equation}
|\tan\theta|< \tan \theta_c := \frac{|\widetilde c-\frac{c}{n-\rho+\kappa c}|}{2\sqrt{-\lambda c}}.
\label{eq:hyperb_local}
\end{equation} 
\label{prop:loc_hyperbolicity}
\end{proposition}

\medskip
\noindent
{\bf Proof:} 
We apply~\cite{frouvelle2011continuum} and find that the hyperbolicity criterion is written:
\begin{equation*}
|\tan\theta|<\frac{|\widetilde c-\frac{\mathrm d}{\mathrm d\rho}(\rho c)|}{2\sqrt{-\lambda c}}.
\end{equation*} 
Using the compatibility condition (\ref{compatibility_condition}) and (\ref{eq:ugly}), eq. (\ref{eq:hyperb_local}) follows. \endproof

As for global hyperbolicity, we have 
\begin{proposition}
We assume (\ref{eq:lambda<0}). Then, system (\ref{rho_eq_z2})-(\ref{v_eq_z2}) is not hyperbolic.
\label{prop:global_hyperbolicity}
\end{proposition}

\medskip
\noindent
{\bf Proof:} It has been proved in~\cite{frouvelle2011continuum} that system~\eqref{mass_conservation_order}-\eqref{orientation_evolution_order} is hyperbolic if and only if~$\lambda>0$. As we assume (\ref{eq:lambda<0}), it follows that the system is not hyperbolic. 
\endproof

We now provide asymptotic expansions of the coefficients which show that, at least when~$\rho \to n$ or~$\rho \to \infty$, we have~$\lambda <0$. 

\begin{proposition} We have the following expansions:

\medskip
\noindent
(i) When~$\rho \to n$:
\begin{align*}
c&=\tfrac{\sqrt{n+2}}{n}\sqrt{\rho-n}+O(\rho-n),\\
\widetilde c&=\tfrac{2n-1}{2n\sqrt{n+2}}\sqrt{\rho-n}+O(\rho-n),\\
\lambda&=\tfrac{-1}{4\sqrt{n+2}}\frac1{\sqrt{\rho-n}}+O(1),\\
\theta_c&=\tfrac{\pi}2-\tfrac{2}{\sqrt{n+2}\sqrt n}\sqrt{\rho-n}+O(\rho-n).
\end{align*}

\medskip
\noindent
(ii) When~$\rho \to \infty$:
\begin{align*}
c&=1- \tfrac{n-1}{2}\rho^{-1} +\tfrac{(n-1)(n+1)}{8}\rho^{-2} + O(\rho^{-3}),\\
\widetilde c&=1-\tfrac{n+1}{2}\rho^{-1}-\tfrac{(n+1)(3n+1)}{24}\rho^{-2}+O(\rho^{-3}),\\
\lambda&= -\tfrac{n+1}{6}\rho^{-2}+O(\rho^{-3}),\\
\theta_c&=\arctan(\tfrac{\sqrt{n+1}\sqrt6}4)+O(\rho^{-1}).
\end{align*}
\label{prop:hyperbolic_expansion}
\end{proposition}

\medskip
\noindent
{\bf Proof:} Using the compatibility condition (\ref{compatibility_condition}), the expression~\eqref{def_lambda2} depends only on~$\kappa$,~$c$, and~$\widetilde c$. With the asymptotic expansion of~$c$ and~$\widetilde c$ as~$\kappa\to0$ and~$\kappa\to\infty$ given in~\cite{frouvelle2011continuum}, we can get an expansion for~$\lambda$.
We have
\begin{align*}
c&=
\begin{cases}
\tfrac1n \kappa-\tfrac1{n^2(n+2)} \kappa^3+O(\kappa^5)&\text{as }\kappa\to0,\\
1- \frac{n-1}{2\kappa} +\frac{(n-1)(n-3)}{8\kappa^2} + O(\kappa^{-3})&\text{as }\kappa\to\infty,
\end{cases}\\
&\\
\widetilde c&=
\begin{cases}
\tfrac{2n-1}{2n(n+2)}\kappa+O(\kappa^2)&\text{as }\kappa\to0,\\
1-\frac{n+1}{2\kappa}+\frac{(n+1)(3n-7)}{24\kappa^2}+O(\kappa^{-3})&\text{as }\kappa\to\infty.
\end{cases}
\end{align*}
We first compute an expansion of~$\rho=\frac{\kappa}c$. We get
\begin{equation}
\rho=\begin{cases}
n+\frac1{n+2}\kappa^2+O(\kappa^4)&\text{as }\kappa\to0,\\
\kappa + \frac{n-1}2 +\frac{(n-1)(n+1)}{8\kappa} + O(\kappa^{-2})&\text{as }\kappa\to\infty.
\end{cases}
\label{exp_rho_kappa}
\end{equation}
Using the definition~\eqref{def_lambda2}, we then get
\begin{equation*}
\lambda=\begin{cases}-\frac1{4\kappa}+O(1)&\text{as } \kappa\to0\\ -\frac{n+1}{6\kappa^2}+O(\kappa^{-3})&\text{as } \kappa\to\infty.\end{cases}
\end{equation*}
We can also expand the threshold angle~$\theta_c$ in terms of~$\kappa$. We get
\begin{equation*}
  \theta_c=
  \begin{cases}\frac{\pi}2-\frac{2}{(n+2)\sqrt n}\kappa+O(\kappa^2)&\text{as }\kappa\to0,\\
    \arctan(\frac{\sqrt{n+1}\sqrt6}4)+O(\kappa^{-1}) &\text{as }\kappa\to\infty.
  \end{cases}
\end{equation*}
We can now reverse the expansion~\eqref{exp_rho_kappa} to get an expansion of~$\kappa$ (and then of the other coefficients) in terms of the density~$\rho$. We get
\begin{equation*}
\kappa=\begin{cases}
\sqrt{n+2}\sqrt{\rho-n}+O(\rho-n)&\text{as }\rho\to n,\\
\rho - \frac{n-1}2 -\frac{(n-1)(n+1)}{8\rho} + O(\rho^{-2})&\text{as }\rho\to\infty.
\end{cases}
\end{equation*}
Inserting this expansion into the previous ones, we finally deduce the expressions stated in proposition \ref{prop:hyperbolic_expansion}. \endproof

When~$\rho \sim n$, since~$|\lambda|=-\lambda$ is large compared to~$\rho$, which is large compared to~$\rho\widetilde c$, the behavior of the orientation equation~\eqref{orientation_evolution_order} can be compared to the behavior of
\begin{equation*}
\partial_t \Omega = \frac{|\lambda|}{\rho} \, (\mathrm{Id} - \Omega \otimes \Omega) \nabla_x \rho,
\end{equation*}
which relaxes~$\Omega$ to the unit vector~$\nabla_x\rho/|\nabla_x\rho|$, with rate
\begin{equation*}
\frac{\lambda}{\rho}|\nabla_x\rho|\sim \tfrac{1}{4n\sqrt{n+2}\sqrt{\rho-n}}|\nabla_x\rho|.
\end{equation*}
This actually makes sense only if the rate of convergence to the equilibrium~$\frac1{\varepsilon}r(\rho)\sim \frac{2{n-1}}{n\varepsilon}(\rho-n)$ in the neighborhood of~$n$ is large compared to this relaxation rate. This requires~$\varepsilon\ll(\rho-n)^{\frac32}|\nabla_x\rho|$. In this case the leading behavior of the system is given by
\begin{equation*}
\partial_t \rho + \nabla_x\cdot\left(\frac{\rho c}{|\nabla_x\rho|} \nabla_x\rho\right)= 0,
\end{equation*}
which is an ill-posed problem, being some kind of nonlinear backwards heat equation. To stabilize this system, a possibility is to derive a first order diffusive correction to model~(\ref{mass_conservation_order}), (\ref{orientation_evolution_order}) using a Chapman-Enskog expansion. Such a correction has been derived in~\cite{degond2010diffusion} for the model of~\cite{degond2008continuum}, but leads to complicated terms. Another possibility is to add some contribution of the non-locality of the interaction in the spirit of \cite{DLMP}. 

\begin{figure}[h]
\begin{minipage}[h]{0.98\linewidth}
\begin{minipage}{0.48\linewidth}
\begin{center}
\begin{picture}(0,0)%
\includegraphics{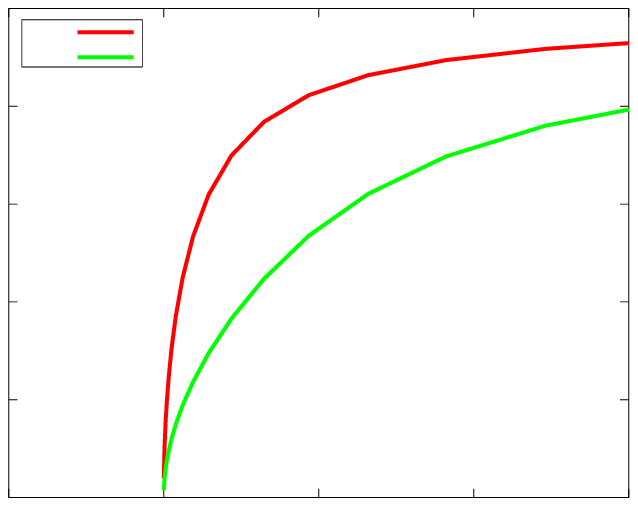}%
\end{picture}%
\setlength{\unitlength}{1579sp}%
\begingroup\makeatletter\ifx\SetFigFont\undefined%
\gdef\SetFigFont#1#2#3#4#5{%
  \reset@font\fontsize{#1}{#2pt}%
  \fontfamily{#3}\fontseries{#4}\fontshape{#5}%
  \selectfont}%
\fi\endgroup%
\begin{picture}(8135,6841)(1852,-7628)
\put(2298,-6894){\makebox(0,0)[rb]{\smash{{\SetFigFont{8}{9.6}{\familydefault}{\mddefault}{\updefault}0}}}}
\put(2298,-5721){\makebox(0,0)[rb]{\smash{{\SetFigFont{8}{9.6}{\familydefault}{\mddefault}{\updefault}0.2}}}}
\put(2298,-4547){\makebox(0,0)[rb]{\smash{{\SetFigFont{8}{9.6}{\familydefault}{\mddefault}{\updefault}0.4}}}}
\put(2298,-3374){\makebox(0,0)[rb]{\smash{{\SetFigFont{8}{9.6}{\familydefault}{\mddefault}{\updefault}0.6}}}}
\put(2298,-2200){\makebox(0,0)[rb]{\smash{{\SetFigFont{8}{9.6}{\familydefault}{\mddefault}{\updefault}0.8}}}}
\put(2298,-1027){\makebox(0,0)[rb]{\smash{{\SetFigFont{8}{9.6}{\familydefault}{\mddefault}{\updefault}1}}}}
\put(2448,-7144){\makebox(0,0)[b]{\smash{{\SetFigFont{8}{9.6}{\familydefault}{\mddefault}{\updefault}0}}}}
\put(4308,-7144){\makebox(0,0)[b]{\smash{{\SetFigFont{8}{9.6}{\familydefault}{\mddefault}{\updefault}2}}}}
\put(6168,-7144){\makebox(0,0)[b]{\smash{{\SetFigFont{8}{9.6}{\familydefault}{\mddefault}{\updefault}4}}}}
\put(8028,-7144){\makebox(0,0)[b]{\smash{{\SetFigFont{8}{9.6}{\familydefault}{\mddefault}{\updefault}6}}}}
\put(9888,-7144){\makebox(0,0)[b]{\smash{{\SetFigFont{8}{9.6}{\familydefault}{\mddefault}{\updefault}8}}}}
\put(6168,-7519){\makebox(0,0)[b]{\smash{{\SetFigFont{8}{9.6}{\familydefault}{\mddefault}{\updefault}Density $\rho$}}}}
\put(3001,-1561){\makebox(0,0)[rb]{\smash{{\SetFigFont{8}{9.6}{\familydefault}{\mddefault}{\updefault}$\widetilde c$}}}}
\put(3001,-1261){\makebox(0,0)[rb]{\smash{{\SetFigFont{8}{9.6}{\familydefault}{\mddefault}{\updefault}$c$}}}}
\end{picture}%
\end{center}
\end{minipage}\hfill
\begin{minipage}{0.48\linewidth}
\begin{center}
\begin{picture}(0,0)%
\includegraphics{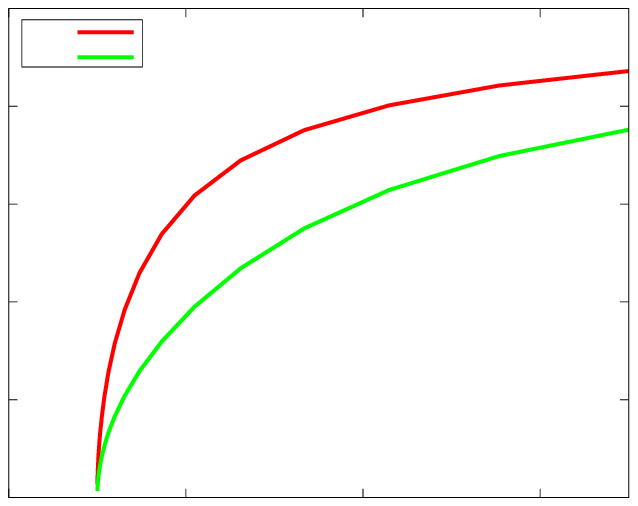}%
\end{picture}%
\setlength{\unitlength}{1579sp}%
\begingroup\makeatletter\ifx\SetFigFont\undefined%
\gdef\SetFigFont#1#2#3#4#5{%
  \reset@font\fontsize{#1}{#2pt}%
  \fontfamily{#3}\fontseries{#4}\fontshape{#5}%
  \selectfont}%
\fi\endgroup%
\begin{picture}(8080,6841)(1852,-7628)
\put(3001,-1561){\makebox(0,0)[rb]{\smash{{\SetFigFont{8}{9.6}{\familydefault}{\mddefault}{\updefault}$\widetilde c$}}}}
\put(3001,-1261){\makebox(0,0)[rb]{\smash{{\SetFigFont{8}{9.6}{\familydefault}{\mddefault}{\updefault}$c$}}}}
\put(2298,-6894){\makebox(0,0)[rb]{\smash{{\SetFigFont{8}{9.6}{\familydefault}{\mddefault}{\updefault}0}}}}
\put(2298,-5721){\makebox(0,0)[rb]{\smash{{\SetFigFont{8}{9.6}{\familydefault}{\mddefault}{\updefault}0.2}}}}
\put(2298,-4547){\makebox(0,0)[rb]{\smash{{\SetFigFont{8}{9.6}{\familydefault}{\mddefault}{\updefault}0.4}}}}
\put(2298,-3374){\makebox(0,0)[rb]{\smash{{\SetFigFont{8}{9.6}{\familydefault}{\mddefault}{\updefault}0.6}}}}
\put(2298,-2200){\makebox(0,0)[rb]{\smash{{\SetFigFont{8}{9.6}{\familydefault}{\mddefault}{\updefault}0.8}}}}
\put(2298,-1027){\makebox(0,0)[rb]{\smash{{\SetFigFont{8}{9.6}{\familydefault}{\mddefault}{\updefault}1}}}}
\put(2448,-7144){\makebox(0,0)[b]{\smash{{\SetFigFont{8}{9.6}{\familydefault}{\mddefault}{\updefault}2}}}}
\put(4574,-7144){\makebox(0,0)[b]{\smash{{\SetFigFont{8}{9.6}{\familydefault}{\mddefault}{\updefault}4}}}}
\put(6699,-7144){\makebox(0,0)[b]{\smash{{\SetFigFont{8}{9.6}{\familydefault}{\mddefault}{\updefault}6}}}}
\put(8825,-7144){\makebox(0,0)[b]{\smash{{\SetFigFont{8}{9.6}{\familydefault}{\mddefault}{\updefault}8}}}}
\put(6168,-7519){\makebox(0,0)[b]{\smash{{\SetFigFont{8}{9.6}{\familydefault}{\mddefault}{\updefault}Density $\rho$}}}}
\end{picture}%
\end{center}
\end{minipage}
\end{minipage}
\caption{The velocities~$c$ and~$\widetilde c$ in dimension~$2$ (left) and~$3$ (right).}
\label{fig_cctild}
\end{figure}

When~$\rho$ is large,~$c$ and~$\widetilde c$ are close to~$1$, and~$\lambda$ is small. In the intermediate regime, numerical computations (see Appendix~2) show that there is a significant difference between~$c$ and~$\widetilde c$. This means that the information about velocity orientation travels slower than the fluid. Fig.~\ref{fig_cctild} displays~$c$ and~$\tilde c$ as functions of~$\rho$, in dimension~$2$ and~$3$.

\begin{figure}[h]
\begin{center}
\begin{picture}(0,0)%
\includegraphics{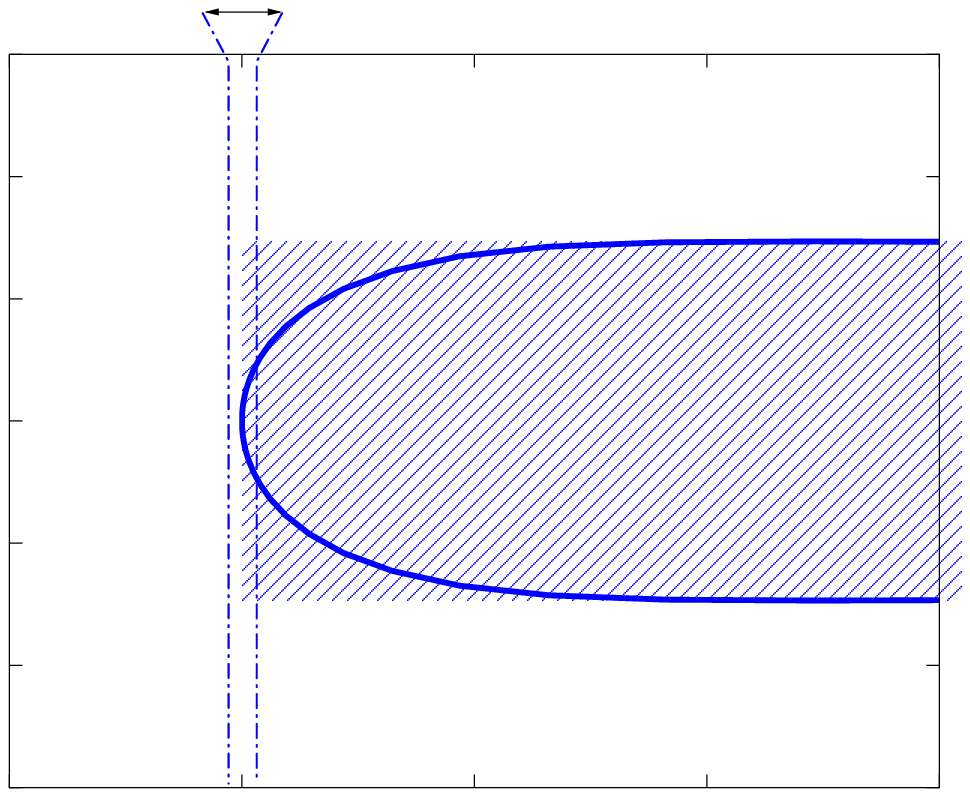}%
\end{picture}%
\setlength{\unitlength}{2368sp}%
\begingroup\makeatletter\ifx\SetFigFont\undefined%
\gdef\SetFigFont#1#2#3#4#5{%
  \reset@font\fontsize{#1}{#2pt}%
  \fontfamily{#3}\fontseries{#4}\fontshape{#5}%
  \selectfont}%
\fi\endgroup%
\begin{picture}(8767,7426)(1220,-7628)
\put(2298,-6894){\makebox(0,0)[rb]{\smash{{\SetFigFont{12}{14.4}{\familydefault}{\mddefault}{\updefault}0}}}}
\put(2298,-5916){\makebox(0,0)[rb]{\smash{{\SetFigFont{12}{14.4}{\familydefault}{\mddefault}{\updefault}$\frac\pi6$}}}}
\put(2298,-4938){\makebox(0,0)[rb]{\smash{{\SetFigFont{12}{14.4}{\familydefault}{\mddefault}{\updefault}$\frac\pi3$}}}}
\put(2298,-3960){\makebox(0,0)[rb]{\smash{{\SetFigFont{12}{14.4}{\familydefault}{\mddefault}{\updefault}$\frac\pi2$}}}}
\put(2298,-2983){\makebox(0,0)[rb]{\smash{{\SetFigFont{12}{14.4}{\familydefault}{\mddefault}{\updefault}$\frac{2\pi}3$}}}}
\put(2298,-2005){\makebox(0,0)[rb]{\smash{{\SetFigFont{12}{14.4}{\familydefault}{\mddefault}{\updefault}$\frac{5\pi}6$}}}}
\put(2298,-1027){\makebox(0,0)[rb]{\smash{{\SetFigFont{12}{14.4}{\familydefault}{\mddefault}{\updefault}$\pi$}}}}
\put(2448,-7144){\makebox(0,0)[b]{\smash{{\SetFigFont{12}{14.4}{\familydefault}{\mddefault}{\updefault}0}}}}
\put(4308,-7144){\makebox(0,0)[b]{\smash{{\SetFigFont{12}{14.4}{\familydefault}{\mddefault}{\updefault}2}}}}
\put(6168,-7144){\makebox(0,0)[b]{\smash{{\SetFigFont{12}{14.4}{\familydefault}{\mddefault}{\updefault}4}}}}
\put(8028,-7144){\makebox(0,0)[b]{\smash{{\SetFigFont{12}{14.4}{\familydefault}{\mddefault}{\updefault}6}}}}
\put(9888,-7144){\makebox(0,0)[b]{\smash{{\SetFigFont{12}{14.4}{\familydefault}{\mddefault}{\updefault}8}}}}
\put(6168,-7519){\makebox(0,0)[b]{\smash{{\SetFigFont{12}{14.4}{\familydefault}{\mddefault}{\updefault}Density $\rho$}}}}
\put(4319,-6149){\makebox(0,0)[b]{\smash{{\SetFigFont{12}{14.4}{\familydefault}{\mddefault}{\updefault}?}}}}
\put(4315,-4808){\makebox(0,0)[b]{\smash{{\SetFigFont{12}{14.4}{\familydefault}{\mddefault}{\updefault}?}}}}
\put(4310,-2977){\makebox(0,0)[b]{\smash{{\SetFigFont{12}{14.4}{\familydefault}{\mddefault}{\updefault}?}}}}
\put(4306,-1770){\makebox(0,0)[b]{\smash{{\SetFigFont{12}{14.4}{\familydefault}{\mddefault}{\updefault}?}}}}
\put(4315,-469){\makebox(0,0)[b]{\smash{{\SetFigFont{12}{14.4}{\familydefault}{\mddefault}{\updefault}$\varepsilon$}}}}
\put(1487,-3772){\rotatebox{90.0}{\makebox(0,0)[b]{\smash{{\SetFigFont{12}{14.4}{\familydefault}{\mddefault}{\updefault}Angle of propagation $\theta$}}}}}
\put(7678,-4009){\makebox(0,0)[b]{\smash{{\SetFigFont{12}{14.4}{\familydefault}{\mddefault}{\updefault}for the reduced model}}}}
\put(7638,-4469){\makebox(0,0)[b]{\smash{{\SetFigFont{12}{14.4}{\familydefault}{\mddefault}{\updefault}\eqref{rho_eq_z2}-\eqref{v_eq_z2}}}}}
\put(7668,-3489){\makebox(0,0)[b]{\smash{{\SetFigFont{12}{14.4}{\familydefault}{\mddefault}{\updefault}Non-hyperbolicity}}}}
\put(7418,-6099){\makebox(0,0)[b]{\smash{{\SetFigFont{12}{14.4}{\familydefault}{\mddefault}{\updefault}Hyperbolicity}}}}
\put(7388,-1699){\makebox(0,0)[b]{\smash{{\SetFigFont{12}{14.4}{\familydefault}{\mddefault}{\updefault}Hyperbolicity}}}}
\put(3487,-3852){\rotatebox{90.0}{\makebox(0,0)[b]{\smash{{\SetFigFont{12}{14.4}{\familydefault}{\mddefault}{\updefault}Non-linear diffusion}}}}}
\put(5167,-3832){\rotatebox{90.0}{\makebox(0,0)[b]{\smash{{\SetFigFont{12}{14.4}{\familydefault}{\mddefault}{\updefault}``Vicsek hydrodynamics''}}}}}
\end{picture}%
\caption{Types of macroscopic limits in dimension~$2$. Around the threshold value~$\rho = 2$, none of the diffusion or hydrodynamic limit is valid. The study of this transition is still open.}
\end{center}
\label{fig_region}.
\end{figure}

Finally, when~$\rho \to \infty$, the critical angle~$\theta_c$ tends to a positive value~$\arctan(\frac{\sqrt{n+1}\sqrt6}4)$. Numerically, we see that~$\theta_c$ is always larger than this limit value. Then, system~\eqref{rho_eq_z2}-\eqref{v_eq_z2} is hyperbolic in the region where the angle~$\theta$ between~$\Omega$ and the direction of propagation is less than this limit value, independently of the density~$\rho$. Fig.~\ref{fig_region} summarizes the different the types of macroscopic limits of the system in dimension~$2$, when the density~$\rho$, and the angle~$\theta$ between~$\Omega$ and the propagation direction vary. The behavior of the system at the crossings either between the hyperbolic and non-hyperbolic regions or between the ordered and disordered regions, remains an open problem. We note that non-hyperbolicity problems appear in other areas such as the motion a an elastic string on a plane~\cite{pego1988instabilities}.  

\setcounter{equation}{0}
\section{Conclusion}
\label{sec:conclu}

In this paper, we have derived a macroscopic model for particles undergoing self-alignment interactions with phase transitions. This model is derived from a time-continuous version of the Vicsek model. We have identified two regimes. In the disordered regime, the macroscopic model is given by a nonlinear diffusion equation depending on the small parameter~$\varepsilon$ describing the ratios of the microscopic to macroscopic length scales. In the ordered regime, the model is given by a hydrodynamic model for self-alignment interaction which is not hyperbolic. Many problems remain open. Among others, a first one is to determine the evolution of the boundary between the ordered and disordered regions and to understand how the models in the two regions are connected across this boundary.  The second one is to understand how to cope with the non-hyperbolicity of the model in the ordered region and possibly modify it by adding small diffusive corrections. Numerical simulations of the particle model are in progress to understand the behavior of the model in the two regimes.

\setcounter{equation}{0}
\section*{Appendix 1. Poincar\'e constant}
\label{appendix_poincare}

In this appendix, we prove the following:

\begin{proposition}
\label{prop_poincare}
We have the following Poincar\'e inequality, for~$\psi\in H^1(\mathbb{S})$:
\begin{equation}
\label{poincare_inequality}
\langle|\nabla_\omega \psi|^2\rangle_{M_{\kappa\Omega}}\geqslant\Lambda_\kappa\langle(\psi-\langle\psi\rangle_{M_{\kappa\Omega}})^2\rangle_{M_{\kappa\Omega}}.
\end{equation}
The best constant~$\Lambda_\kappa$ in this inequality is the smallest positive eigenvalue of the operator
\begin{equation}
L_{\kappa\Omega}^*=-\frac1{M_{\kappa\Omega}}\nabla_\omega\cdot ( M_{\kappa\Omega}\nabla_\omega \cdot ).
\label{eq:def_LkO*} 
\end{equation}

We define the linear operator~$L_\kappa^*$ by
\begin{equation}
L_\kappa^*(g)(\theta)=-(\sin \theta)^{2-n} e^{-\kappa\cos\theta}((\sin \theta)^{n-2} e^{\kappa\cos\theta}g'(\theta))'.
\label{def_Lstar_SL}
\end{equation}
Then one of the following three possibilities is true:

\medskip
\noindent
(i)~$\Lambda_\kappa$ is the smallest eigenvalue of the Sturm-Liouville problem
\begin{equation}
L_\kappa^*(g)=\lambda g,
\label{eq:Y}
\end{equation}
for~$g\in C^2([0,\pi])$ with Neumann boundary conditions ($g'(0)=g'(\pi)=0$) and such that~$\int_0^\pi(\sin \theta)^{n-2} e^{\kappa\cos\theta}g(\theta)\mathrm d\theta=0$, and the eigenspace of~$L_{\kappa\Omega}^*$ associated to the eigenvalue~$\Lambda_\kappa$ is of dimension~$1$, spanned by~$\omega\mapsto h^0_\kappa(\omega\cdot\Omega)$, where the function~$\theta\mapsto h_0(\cos\theta)$ is smooth, positive for~$0\leqslant\theta<\theta_0$ and negative for~$\theta_0<\theta\leqslant\pi$.

\medskip
\noindent
(ii)~$\Lambda_\kappa$ is the smallest eigenvalue of the Sturm-Liouville problem
\begin{equation}
\widetilde L_\kappa^*(g)=L_\kappa^*(g)+\tfrac{n-2}{\sin^2\theta}g(\theta)=\lambda g,
\label{eq:X}
\end{equation}
for~$g\in C^2([0,\pi])$ with Dirichlet boundary conditions ($g(0)=g(\pi)=0$), and the eigenspace of~$L_{\kappa\Omega}^*$ associated to~$\Lambda_\kappa$ is of dimension~$n-1$, consisting in the functions of the form~$\psi_A(\omega)=h_\kappa^1(\omega\cdot\Omega)A\cdot\omega$ for any vector~$A\in\mathbb{R}^n$ such that~$\Omega\cdot A=0$, with~$\theta\mapsto h_\kappa^1(\cos\theta)$ a smooth positive function for~$0<\theta<\pi$.

\medskip
\noindent
(iii) The two above Sturm-Liouville problems have the same smallest eigenvalue~$\Lambda_\kappa$, and the eigenspace of~$L_{\kappa\Omega}^*$ associated to~$\Lambda_\kappa$ is of dimension~$n$, spanned by the two types of function of the above cases.
\end{proposition} 

\begin{proof}
First of all, we have
\begin{equation}
\langle|\nabla_\omega \psi|^2\rangle_{M_{\kappa\Omega}} \, \geqslant\, (\min M_{\kappa\Omega}) \, \int_\mathbb{S} |\nabla_\omega\psi|^2 \, \geqslant \, (\min M_{\kappa\Omega}) \, (n-1) \, \int_\mathbb{S} \left(\psi-\int_\mathbb{S} \psi \right)^2,
\label{eq:ugly_A}
\end{equation}
and
\begin{equation}
\langle(\psi-\langle\psi\rangle_{M_{\kappa\Omega}})^2\rangle_{M_{\kappa\Omega}}\, \leqslant \, \langle(\psi-\int_\mathbb{S} \psi)^2\rangle_{M_{\kappa\Omega}} \, \leqslant \, (\max M_{\kappa\Omega}) \, \int_\mathbb{S} \left(\psi- \int_\mathbb{S} \psi \right)^2.
\label{eq:ugly_B}
\end{equation}
The second inequality of (\ref{eq:ugly_A}) follows from the Poincar\'e inequality on the sphere:
$$ \int_\mathbb{S} \left(\psi-\int_\mathbb{S} \psi \right)^2 \leqslant \frac{1}{n-1} \int_\mathbb{S} |\nabla_\omega\psi|^2.~$$
The first inequality of (\ref{eq:ugly_B}) follows from the fact that 
$$ \langle(\psi-\int_\mathbb{S} \psi)^2\rangle_{M_{\kappa\Omega}} - \langle(\psi-\langle\psi\rangle_{M_{\kappa\Omega}})^2\rangle_{M_{\kappa\Omega}} = \left( \int_\mathbb{S} \psi - \int_\mathbb{S} \psi M_{\kappa\Omega} \right) ^2 \geqslant0 .$$
Eqs. (\ref{eq:ugly_A}) and (\ref{eq:ugly_B}) lead to the Poincar\'e inequality~\eqref{poincare_inequality} with
$$ \Lambda_\kappa\, \geqslant \, (n-1) \, \frac{\min M_{\kappa\Omega}}{\max M_{\kappa\Omega}} \, = \, (n-1)\,e^{2\kappa} .~$$

We use the inner product~$(\varphi,\psi)\mapsto\langle\varphi\psi\rangle_{M_{\kappa\Omega}}$, adapted to~$M_{\kappa\Omega}$. 
We denote by~$\dot{L}^2_\kappa(\mathbb{S})$ (resp.~$\dot{H}^1_\kappa(\mathbb{S})$) the functions~$\psi\in L^2(\mathbb{S})$ (resp. in~$H^1(\mathbb{S})$) such that~$\langle \psi \rangle_{M_{\kappa\Omega}}=0$.

The operator~$L_{\kappa\Omega}^*$ given by (\ref{eq:def_LkO*}) is self-adjoint since~$\langle\nabla_\omega \psi\cdot\nabla_\omega\varphi\rangle_{M_{\kappa\Omega}}=\langle\psi L_{\kappa\Omega}^*\varphi\rangle_{M_{\kappa\Omega}}$. It is then easy to see, using Lax-Milgram theorem, that if~$\varphi$ belongs to~$\dot{L}^2_\kappa(\mathbb{S})$ then there is a unique solution~$\psi\in\dot{H}^1_\kappa(\mathbb{S})$ to the equation~$L_{\kappa\Omega}^*\psi=\varphi$. The so-obtained inverse operator is then compact and self-adjoint. By the spectral theorem, we get a basis of eigenfunctions, in the Hilbert space~$\dot{L}^2_\kappa(\mathbb{S})$, which are also eigenfunctions of~$L_{\kappa\Omega}^*$. If we denote~$\Lambda_\kappa^{-1}$ the largest eigenvalue of the inverse of~$L_{\kappa\Omega}^*$, then it is easy to see that~$\Lambda_\kappa$ is the best constant for the following Poincar\'e inequality, in the space~$\dot{H}^1_\kappa(\mathbb{S})$:
\begin{equation*}
\langle|\nabla_\omega \psi|^2\rangle_{M_{\kappa\Omega}} \, \geqslant \, 
\Lambda_\kappa \, \langle\psi^2\rangle_{M_{\kappa\Omega}} \, \geqslant \,
\Lambda_\kappa \, \langle (\psi - \langle \psi \rangle_{M_{\kappa\Omega}} )^2 \rangle_{M_{\kappa\Omega}}
.
\end{equation*}
Since the constants trivially satisfy this inequality, this shows that~$\Lambda_\kappa$ is the best constant for the Poincar\'e inequality~\eqref{poincare_inequality} in~$H^1(\mathbb{S})$.

The goal is now to reduce the computation of the eigenvalues to simpler problems, using separation of variables. We write~$\omega=\cos\theta\,\Omega+\sin\theta\,v$, where~$v$ belongs to the unit sphere, orthogonal to~$\Omega$. We identify~$\Omega$ with the last element of an orthogonal basis of~$\mathbb{R}^n$, and we write~$v\in\mathbb{S}_{n-2}$.

By spherical harmonic decomposition in an adapted basis (see for example~\cite{frouvelle2011dynamics}, appendix~A), we have a unique decomposition of the form
\begin{equation}
\psi(\omega)={\sum_{k,m}} g_{m}^k(\theta)Z_{m}^k(v),
\label{adapted_decomposition}
\end{equation}
where~$(Z_{m}^k(v))_{k\in\llbracket1,k_m\rrbracket}$ is a given orthonormal basis of the spherical harmonics of degree~$m$ on~$\mathbb{S}_{n-2}$, for~$m\in\mathbb{N}$, with~$k_m=\tbinom{n+m-2}{n-2}- \tbinom{n+m-4}{n-2}$. If~$\psi$ is continuous,~$g_{m}^k$ is given by
\begin{equation}
g_{m}^k(\theta)=\int_{\mathbb{S}_{n-2}}\psi(\cos\theta \, \Omega+\sin\theta \, v)Z_{m}^k(v)\mathrm d v.
\label{def_gmk}
\end{equation}
We now show that the decomposition (\ref{adapted_decomposition}) remains stable under the action of the operator~$L_{\kappa \Omega}$, so that its spectral decomposition can be performed independently for each term of the decomposition.

First, we examine the case of dimension~$n\geqslant3$. Let~$\psi(\omega)=g(\theta)Z(v)$. We have 
\begin{equation*}
\nabla_\omega\psi(\omega)=g'(\theta) e_\theta Z(v) + \frac{g(\theta)}{\sin\theta}\nabla_vZ(v),
\end{equation*}
where the unit vector~$e_\theta$ is given by 
$$ e_\theta = \nabla_\omega\theta=-\frac1{\sin\theta}(\mbox{Id}-\omega\otimes\omega)\Omega.$$ 

We take functions~$\psi(\omega)= g_{m}^k(\theta)Z_{m}^k(v)$ and~$\varphi(\omega)=\sum_{k,m} f_{m}^k(\theta)Z_{m}^k(v)$. Since the spherical harmonics are orthonormal, and are eigenfunctions of~$\Delta_v$ for the eigenvalues~$-m(m+n-3)$, we get:
\begin{equation}
\langle\nabla_\omega \psi\cdot\nabla_\omega\varphi\rangle_{M_{\kappa\Omega}}=\int_0^\pi[{f_{m}^k}'(\theta){g_{m}^k}'(\theta)+\tfrac{m(m+n-3)}{\sin^2\theta}f_{m}^k(\theta)g_{m}^k(\theta)](\sin \theta)^{n-2} e^{\kappa\cos\theta}\mathrm d\theta.
\label{eq:ugly_AA}
\end{equation}

Suppose~$m\geqslant1$. Then, it is easy to see that the function~$\psi$ belongs to~$\dot{H}^1_\kappa(\mathbb{S})$ if and only if~$(\sin \theta)^{\frac{n}2-1} g'\in L^2(0,\pi)$ and~$(\sin \theta)^{\frac{n}2-2} g'\in L^2(0,\pi)$. This condition is equivalent to the fact that~$g\in V$, where~$V$ is defined by~\eqref{def_V2}, and which we denote by~$V_\kappa^m$ for convenience:
\begin{equation*}
V_\kappa^m = \{ g \, | \,(\sin\theta)^{\frac n2-2} g \in L^2(0,\pi), \, (\sin\theta)^{\frac n2-1}g \in H^1_0(0,\pi) \}.
\end{equation*}
Suppose now that~$m=0$. Then~$Z^k_m$ is a constant, and the condition~$\psi\in\dot{H}^1_\kappa(\mathbb{S})$ is equivalent to the first condition only:~$(\sin \theta)^{\frac{n}2-1} g'\in L^2(0,\pi)$, under the constraint that~$\int_0^\pi(\sin \theta)^{n-2} e^{\kappa\cos\theta}g(\theta)\mathrm d\theta=0$. We will denote this space by~$V^0_\kappa$:
\begin{equation*}
V_\kappa^0 = \{ g \, | \,(\sin\theta)^{\frac n2-1} g' \in L^2(0,\pi), \, \textstyle\int_0^\pi(\sin \theta)^{n-2} e^{\kappa\cos\theta}g(\theta)\mathrm d\theta=0 \}.
\end{equation*}
Formula (\ref{eq:ugly_AA}) then suggests to define the operator~$L_{\kappa,m}^*:V_\kappa^m\to(V_\kappa^m)^*$ by 
\begin{equation}
\int_0^\pi f(\theta) \, L_{\kappa,m}^*g(\theta) \, (\sin \theta)^{n-2} \, e^{\kappa\cos\theta} \, \mathrm d\theta \, =\, \int_0^\pi [f'g' \, + \, \tfrac{m(m+n-3)}{\sin^2\theta}fg] \, (\sin \theta)^{n-2} \,  e^{\kappa\cos\theta}\, \mathrm d\theta.
\label{def_Lkm}
\end{equation}
From (\ref{eq:ugly_AA}), it follows that, if we decompose~$\psi(\omega)=\sum_{k,m} g_{m}^k(\theta)Z_{m}^k(v)$, then 
\begin{equation*}
L_{\kappa\Omega}^*\psi(\omega)=\sum_{k,m} L_{\kappa,m}^*g_{m}^k(\theta) \, Z_{m}^k(v),
\end{equation*}
showing that~$L_{\kappa\Omega}^*$ is block diagonal on each of these spaces~$V_\kappa^m$ (tensorized by the spherical harmonics of degree~$m$ on~${\mathbb S}_{n-2}$). 
So we can perform the spectral decomposition of~$L_{\kappa\Omega}^*$ by means of the spectral decomposition of each of the~$L_{\kappa,m}^*$. It is indeed easy to prove, using Lax-Milgram theorem, that the operators~$L_{\kappa,m}^*$ have self-adjoint compact inverses for the dot product~$(f,g)=\int_0^\pi fg(\sin \theta)^{n-2} e^{\kappa\cos\theta}\mathrm d\theta$. Therefore the eigenfunctions and eigenvalues of~$L_{\kappa\Omega}^*$ correspond to those of the operators~$L_{\kappa,m}^*$, for all~$m\in\mathbb{N}$.
If we denote by~$\lambda_{\kappa,m}$ the smallest eigenvalue of~$L_{\kappa,m}^*$, we finally get 
\begin{equation*}
\Lambda_\kappa=\min\{\lambda_{\kappa,m},m\in\mathbb{N}\}.
\end{equation*}
We notice that
\begin{eqnarray*}
& & \hspace{-1cm} 
\lambda_{\kappa,m}=\inf \left\{ \left. \int_0^\pi f(\theta) \, L_{\kappa,m}^*f(\theta) \, (\sin \theta)^{n-2} \,  e^{\kappa\cos\theta} \, \mathrm d\theta \quad \right| \quad f\in V_\kappa^m, \right. \\
& & \hspace{7cm} \left. 
\int_0^\pi f^2(\theta) \, (\sin \theta)^{n-2} \, e^{\kappa\cos\theta} \, \mathrm d\theta=1 \right\} ,
\end{eqnarray*}
but since all the~$V_\kappa^m$ are the same for~$m\geqslant1$, and since
\begin{equation*}
\int_0^\pi \tfrac{1}{\sin^2\theta} \, f^2  \, (\sin \theta)^{n-2} \, e^{\kappa\cos\theta} \, \mathrm d\theta \, \geqslant \, \int_0^\pi f^2 \, (\sin \theta)^{n-2} \, e^{\kappa\cos\theta} \, \mathrm d\theta,
\end{equation*}
we get
\begin{equation*}
\lambda_{\kappa,m+1}\geqslant\lambda_{\kappa,m}+(m+1)(m+n-2)-m(m+n-3)=\lambda_{\kappa,m}+2m+n-2.
\end{equation*}
Finally,~$\Lambda_\kappa$ is the minimum between~$\lambda_{\kappa,0}$ and~$\lambda_{\kappa,1}$. The eigenfunctions for the operator~$L_{\kappa\Omega}^*$ being smooth, this is also true for the operators~$L_{\kappa,m}^*$, by formula~\eqref{def_gmk}. So we can transform the definitions~\eqref{def_Lkm} by integration by parts. 

Indeed, if~$g_0$ is an eigenfunction (in~$V_\kappa^0$) associated to~$L_{\kappa,0}^*$ and an eigenvalue~$\lambda$, then~$g_0$ is smooth and satisfies the Sturm-Liouville eigenvalue problem
\begin{equation*}
L_\kappa^*g_0(\theta)=-(\sin \theta)^{2-n} e^{-\kappa\cos\theta}((\sin \theta)^{n-2} e^{\kappa\cos\theta}g'(\theta))'=\lambda g_0(\theta).
\end{equation*}
Conversely, a smooth function with the condition~$\int_0^\pi(\sin \theta)^{n-2} e^{\kappa\cos\theta}g(\theta)\mathrm d\theta=0$ belongs to~$V_\kappa^0$. Actually, in dimension~$n\geqslant3$, we do not need to impose the Neumann boundary conditions: they appear naturally, since we have 
\begin{equation*}
L_\kappa^*g_0=-e^{-\kappa\cos\theta}(e^{\kappa\cos\theta}g_0')'-\tfrac{n-2}{\tan\theta} g_0'=\lambda g_0.
\end{equation*}
Therefore by continuity at~$\theta=0$ and~$\pi$,~$g_0'(0)=g_0'(\pi)=0$.
Then, using classical Sturm-Liouville oscillation theory (see \cite{weidmann1987spectral} for example), we find that the first eigenspace of~$L_\kappa^*$ is of dimension~$1$, spanned by a function~$g_{\kappa,0}(\theta)$, which is positive for~$0\leqslant\theta<\theta_0$ and negative for~$\theta_0<\theta\leqslant\pi$.

Similarly, if~$g_1$ is an eigenfunction (in~$V_\kappa^1$) associated to~$L_{\kappa,1}^*$ and an eigenvalue~$\lambda$,then~$g_1$ is smooth, with~$g_1(0)=g_1(\pi)=0$ and satisfies the Sturm-Liouville eigenvalue problem
\begin{equation*}
\widetilde L_{\kappa,1}^*g_1(\theta)=L_\kappa^*g_1(\theta)+\tfrac{n-2}{\sin^2\theta}g_1(\theta)=\lambda g_1(\theta).
\end{equation*}
And conversely, if a function satisfies Dirichlet boundary conditions while being in~$C^2([0,\pi])$, it belongs to~$V_\kappa^1$. Once again, if~$n \geqslant 3$, we do not need to impose the Dirichlet boundary conditions in the~$C^2([0,\pi])$ framework, since we have 
\begin{equation*}
L_\kappa^*g_1=-e^{-\kappa\cos\theta}(e^{\kappa\cos\theta}g_1')'-\tfrac{n-2}{\tan\theta} g_1'+\tfrac{n-2}{\sin^2\theta}g_1=\lambda g_1.
\end{equation*}
So, by continuity at~$\theta=0$ and~$\pi$,~$g_1(0)=g_1(\pi)=0$, and then a first order expansion shows that continuity holds whatever the values of~$g_0'(\theta)$ at the endpoints are. Again, using classical Sturm-Liouville theory, we find that the first eigenspace of~$L_\kappa^*$ is of dimension~$1$, spanned by a function~$g_{\kappa,1}(\theta)$, which keeps the same sign on~$(0,\pi)$.

The case~$\lambda_{\kappa,0}<\lambda_{\kappa,1}$ corresponds to case (i) of the proposition. Since a spherical harmonic of degree~$0$ on the sphere~$\mathbb{S}_{n-2}$ is a constant, introducing~$h_\kappa^0$ such that ~$h_\kappa^0(\cos\theta)=g_{\kappa,0}(\theta)$ allows us to state that the eigenspace of~$L_{\kappa\Omega}^*$ associated to the lowest eigenvalue is spanned by~$\omega\mapsto h_\kappa^0(\omega\cdot\Omega)$. 

The case~$\lambda_{\kappa,0}>\lambda_{\kappa,1}$ corresponds to case (ii) of the proposition. The spherical harmonics of degree~$1$ on the sphere~$\mathbb{S}_{n-2}$ are the functions of the form~$v\mapsto A\cdot v$, with~$A\cdot\Omega=0$. Introducing~$h_\kappa^0$ such that ~$h_\kappa^0(\cos\theta) \, \sin\theta = g_{\kappa,0}(\theta)$ allows us to state that the eigenspace of~$L_{\kappa\Omega}^*$ associated to the lowest eigenvalue is of dimension~$n-1$, consisting of the functions of the form~$\omega\mapsto h_\kappa^1(\omega\cdot\Omega)\, A\cdot\omega$, with~$A$ any vector in~$\mathbb{R}^n$ such that~$A \cdot \Omega =0$. 

Finally, the case~$\lambda_{\kappa,0}=\lambda_{\kappa,1}$ corresponds to case (iii) of the proposition and this ends the proof in the case of dimension~$n\geqslant3$. 

We now examine the special case of dimension~$n=2$. We identify~$H^1(\mathbb{S})$ with the~$2\pi$-periodic functions in~$H^1_{loc}(\mathbb{R})$. So,~$\Lambda_\kappa$ is the smallest eigenvalue of the periodic Sturm-Liouville problem
\begin{equation*}
L_\kappa^*(g)=\widetilde L_\kappa^*(g)=-e^{-\kappa\cos\theta}(e^{\kappa\cos\theta}g')'=\lambda g,
\end{equation*}
for functions~$g$ such that~$\int_{-\pi}^\pi e^{\kappa\cos\theta}g(\theta)\mathrm d\theta=0$.
Here the decomposition corresponding to~\eqref{adapted_decomposition} is the even-odd decomposition (there are only two spherical harmonics on~$\mathbb{S}_0$: the constant function of degree~$0$ and the odd function of degree~$1$).
The odd part~$g_o$ of~$g$ can be identified with a function of~$H^1_0(0,\pi)$, and it is easy to see that the odd part of~$L_\kappa^*(g)$ is~$L_\kappa^*(g_o)$, and similarly for the even part~$g_e$.
So, we can perform the spectral decomposition of~$L^*_\kappa$ separately on the spaces of even and odd functions.

Actually, if~$g$ is a solution of the Sturm-Liouville periodic problem, the function~$\widetilde g(\theta)=e^{-\kappa\cos\theta}\partial_\theta g(\pi-\theta)$ is another solution with the same eigenvalue. Furthermore, if~$g$ is odd, then~$\widetilde g$ is even and conversely. So the eigenvalues are the same for the odd and even spaces problems. Therefore, in dimension~$n=2$, proposition \ref{prop_poincare} can be refined and we can state that case (iii) is the only possibility: the eigenspace of~$L_{\kappa\Omega}^*$ associated to~$\Lambda_\kappa$ is of dimension~$2$, spanned by an odd function~$g_\kappa^o$, positive on~$(0,\pi)$, and an even function~$g_\kappa^e=\widetilde g_\kappa^o$, positive for~$0<\theta<\theta_0$ and negative for~$\theta_0<\theta<\pi$. The proof of Proposition \ref{prop_poincare} is complete. 
\end{proof}

We can now state a conjecture, which refines proposition \ref{prop_poincare}, if true, and which is based on numerical experiments. 

\begin{conjecture}
(i) When~$\kappa>0$ and~$n\geqslant3$, only statement (ii) of Proposition~\ref{prop_poincare} is true. 

\medskip
\noindent
(ii) The function~$\kappa\mapsto\Lambda_\kappa$ is increasing.

\label{conj:poincare}
\end{conjecture}

We also observe numerically that~$\lambda_1\sim \kappa$ when kappa is large.

Some investigations are in progress to prove the monotonicity of the eigenvalue with respect to~$\kappa$, based on formal expansions similar to those used in Section~$5$ of~\cite{frouvelle2011continuum}.
 
\begin{remark}
 At the end of the proof of Proposition~\ref{prop_poincare}, we have seen that in dimension~$n=2$ only statement (iii) is true. The proof uses a transformation of the solution of an eigenvalue problem into the solution of another eigenvalue problem. We can try to find a similar transformation in dimensions~$n\geqslant3$: if~$f$ satisfies~$L_{\kappa,0}f=\lambda f$ (with Neumann boundary conditions) then~$\widetilde f=e^{-\kappa\cos\theta}\, \partial_\theta f(\pi-\theta)$ (with Dirichlet boundary conditions) satisfies
\begin{equation*}
\begin{split}
\int_0^\pi \widetilde f \, L_1\widetilde f \, (\sin \theta)^{n-2} \, e^{\kappa\cos\theta} \, \mathrm d\theta&=\lambda\int_0^\pi \widetilde f^2 \, (\sin \theta)^{n-2} \,  e^{\kappa\cos\theta} \, \mathrm d\theta\\
&-\kappa(n-2) \, \int_0^\pi \cos\theta \, \widetilde f^2 \, (\sin \theta)^{n-2} \,  e^{\kappa\cos\theta} \, \mathrm d\theta,
\end{split}
\end{equation*}
so if we can prove that~$\int_0^\pi \cos\theta\, \widetilde f^2 \, (\sin \theta)^{n-2} \,  e^{\kappa\cos\theta} \, \mathrm d\theta>0$, we can deduce that~$\lambda_0>\lambda_1$. So far we have been unable to prove this estimate.
\end{remark}

\setcounter{equation}{0}
\section*{Appendix 2. Numerical computations of the coefficients}
\label{appendix_numerics}

We adopt a finite difference approach to compute the function ~$g_\kappa$ associated to the GCI's and defined by (\ref{def_GCI_elliptic}).
We consider the function~$f_\kappa$ such that~$f_\kappa(\theta)=(\sin \theta)^{\frac{n}2-1} g_\kappa(\theta)$. In particular, since~$g_\kappa\in V$ defined by~\eqref{def_V2},~$f_\kappa$ belongs to~$H^1_0(0,\pi)$. Since~$g_\kappa$ satisfies (\ref{def_GCI_elliptic}),~$f_\kappa$ satisfies 
\begin{equation*}
-e^{-\kappa\cos\theta}(e^{\kappa\cos\theta}f_\kappa')'+(\tfrac{n-2}{2\sin^2\theta}(1+\tfrac{n-2}2\cos^2\theta)-\kappa\cos\theta)f_\kappa=\sin^{\frac{n}2}\theta.
\end{equation*}
We discretize the interval~$(0,\pi)$ with~$N+1$ points~$\theta_i=\tfrac1Ni\pi$, and denote by~$f^i_\kappa$ an approximation of~$f_\kappa$ at these points. Since~$f_\kappa\in H^1_0(0,\pi)$,~$f^0_\kappa=f^N_\kappa=0$. We define~$e_\kappa^i=e^{\kappa\cos\theta_i}$. A second order approximation of~$(e^{\kappa\cos\theta}f_\kappa')'$ at~$\theta_i$ is then given by
\begin{equation*}
(e^{\kappa\cos\theta}f_\kappa')' (\theta_i) \approx \frac{N^2}{\pi^2}(e_\kappa^{i+\frac12}(f^{i+1}_\kappa-f^{i}_\kappa)-e_\kappa^{i-\frac12}(f^{i}_\kappa-f^{i-1}_\kappa)).
\end{equation*}
Introducing~
\begin{gather*}
d_\kappa^i=\frac{n-2}{2\sin^2\theta_i}(1+\tfrac{n-2}2\cos^2\theta_i)-\kappa\cos\theta_i+\frac{N^2}{\pi^2}\frac{e_\kappa^{i-\frac12}+e_\kappa^{i+\frac12}}{e_k^i},\\
b_\kappa^i=-\frac{N^2}{\pi^2}\frac{e_\kappa^{i+\frac12}}{e_k^i},\quad\text{ and }\quad \widetilde b_\kappa^i=-\frac{N^2}{\pi^2}\frac{e_\kappa^{i-\frac12}}{e_k^i},
\end{gather*}
the vector~$F=(f^i_\kappa)_{i\in\llbracket1,N-1\rrbracket}$ is the solution of the linear system~$AF=S$, where the vector~$S$ is~$(\sin^{\frac{n}2}\theta_i)_{i\in\llbracket1,N-1\rrbracket}$, and the tridiagonal matrix~$A$ is defined by
\begin{equation}
\label{def_A}
A=
\begin{pmatrix}
d_\kappa^1            & b_\kappa^1 &    0   & \dots  & \dots & 0\\
\widetilde b_\kappa^2 & d_\kappa^2 & b_\kappa^2  & \ddots &   & \vdots \\
0 & \widetilde b_\kappa^3 & d_\kappa^3 & \ddots & \ddots  &  \vdots \\
\vdots & \ddots           & \ddots        & \ddots & b_\kappa^{N-3}   &0\\
\vdots  & & \ddots& \widetilde b_\kappa^{N-2}& d_\kappa^{N-2} & b_\kappa^{N-2}\\
0  &\dots &\dots &0 & \widetilde b_\kappa^{N-1} & d_\kappa^{N-1}
\end{pmatrix}.
\end{equation}
We use the trapezoidal method to perform the integrations in the definitions~\eqref{def_c} and \eqref{def_ctild} of~$c$ and~$\widetilde c$. The other coefficients~$\rho$,~$\lambda$ and~$\theta_c$ are then directly computed from~$c$ and~$\widetilde c$.
The numerical results provided in Figures~\ref{fig_cctild}-\ref{fig_region} have been obtained for~$N=3000$.

We now detail how we obtain an approximation of the Poincar\'e constant~$\Lambda_\kappa$. By Appendix~1,~$\Lambda_\kappa$ is the minimum between~$\lambda_{\kappa,1}$ and~$\lambda_{\kappa,0}$, which are the smallest eigenvalue of two Sturm-Liouville problems. Several algorithms exist to compute eigenvalues of singular Sturm-Liouville problems (which is the case here whenever~$n\geqslant3$) with a good precision~\cite{bailey1991computing}. However, we use a simpler method based on finite differences.

Actually,~$\lambda_{\kappa,1}$ is the smallest eigenvalue associated to problem (\ref{eq:X}), with~$g\in V$. So, considering once again the function~$f$ such that~$f(\theta)=(\sin \theta)^{\frac{n}2-1} g(\theta)$, the vector~$AF$, with~$A$ defined by~\eqref{def_A}, gives a second order approximation of~$(\sin \theta)^{\frac{n}2-1}\widetilde L_\kappa^*g(\theta)=\lambda f(\theta)$ at the points~$\theta_i$. So we can take the smallest eigenvalue of~$A$ as an approximation of~$\lambda_{\kappa,1}$.

We now look for an approximation of~$\lambda_{\kappa,0}$. Let~$g$ be a solution of the Sturm-Liouville problem (\ref{eq:Y}) with Neumann boundary conditions. We introduce~$G=(g_{i+\frac12})_{i\in\llbracket0,N-1\rrbracket}$, the vector of approximations of~$g$ at the points~$\theta_{i+\frac12}=\tfrac1N(i+\frac12)\pi$. Introducing~$m_\kappa^i=(\sin \theta)^{n-2} \, e^{\kappa\cos\theta_i}$, a second order approximation of~$L_\kappa^*g$ at the point~$\theta_{i+\frac12}$, with~$i\in\llbracket1,N-2\rrbracket$ is then given by
\begin{equation*}
L_\kappa^*g (\theta_{i+\frac12}) \approx
\frac{N^2}{\pi^2m_\kappa^{i+\frac12}}(-m_\kappa^{i+1}(f^{i+\frac32}_\kappa-f^{i+\frac12}_\kappa)+m_\kappa^{i}(f^{i+\frac12}_\kappa-f^{i-\frac12}_\kappa)).
\end{equation*}
With the Neumann boundary conditions, the approximations at the points~$\theta_{\frac12}$ and~$\theta_{N-\frac12}$ are given by
\begin{equation*}
L_\kappa^*g (\theta_{\frac12}) \approx \frac{N^2}{\pi^2m_\kappa^{\frac12}}m_\kappa^{1}(f^{\frac32}_\kappa-f^{\frac12}_\kappa), \quad \quad L_\kappa^*g (\theta_{N-\frac12}) \approx -\frac{N^2}{\pi^2m_\kappa^{N-\frac12}}m_\kappa^{N-1}(f^{N-\frac12}_\kappa-f^{N-\frac32}_\kappa).
\end{equation*}
Introducing
\begin{gather*}
d_\kappa^{i+\frac12}=\frac{N^2}{\pi^2}\frac{m_\kappa^{i+1}+m_\kappa^{i}}{m_k^{i+\frac12}},\\
b_\kappa^{i+\frac12}=-\frac{N^2}{\pi^2}\frac{m_\kappa^{i+1}}{m_k^{i+\frac12}},\quad\text{ and }\quad \widetilde b_\kappa^{i+\frac12}=-\frac{N^2}{\pi^2}\frac{m_\kappa^{i}}{m_k^{i-\frac12}},
\end{gather*}
a second order approximation of~$L_\kappa^*g$ is given by~$BG$, where the tridiagonal matrix~$B$ is defined by
\begin{equation}
B=
\begin{pmatrix}
-b_\kappa^{\frac12}            & b_\kappa^{\frac12} &    0   & \dots  & \dots & 0\\
\widetilde b_\kappa^{\frac32} & d_\kappa^{\frac32} & b_\kappa^{\frac32}  & \ddots &   & \vdots \\
0 & \widetilde b_\kappa^{\frac52} & d_\kappa^{\frac52} & \ddots & \ddots  &  \vdots \\
\vdots & \ddots           & \ddots        & \ddots & b_\kappa^{N-\frac52}   &0\\
\vdots  & & \ddots& \widetilde b_\kappa^{N-\frac32}& d_\kappa^{N-\frac32} & b_\kappa^{N-\frac32}\\
0  &\dots &\dots &0 & \widetilde b_\kappa^{N-\frac12} &  -\widetilde b_\kappa^{N-\frac12}
\end{pmatrix},
\end{equation}

So we can take the smallest positive eigenvalue of~$B$ as an approximation of~$\lambda_{\kappa,0}$ (excluding the constant functions).
The computations of Fig.~\ref{fig_rates} have been performed with~$N=300$ points.


\end{document}